\documentclass[11pt,a4paper]{article}
\usepackage{jheppub}
\def\beq{\begin{equation}}
\def\eeq{\end{equation}}

\def\beq{\begin{equation}}
\def\eeq{\end{equation}}
\def\bea{\begin{eqnarray}}
\def\eea{\end{eqnarray}}

\def\fig#1{{Fig.~\ref{#1}}}

\parskip=\itemsep               %?

\def\fig#1{{Fig.~\ref{#1}}}
\title{ The JIMWLK evolution and the s-channel unitarity.}
\author[a]{Alex Kovner,}
\author[b,c]{Eugene Levin,}
\author[a]{Ming Li}
\author[d,a]{and Michael Lublinsky}
\affiliation[a]{Physics Department, University of Connecticut, 2152 Hillside Road, Storrs, CT 06269, USA}
\affiliation[b]{Departemento de F\'isica, Universidad T\'ecnica Federico Santa Mar\'ia, and Centro Cient\'ifico-\\
Tecnol\'ogico de Valpara\'iso, Avda. Espana 1680, Casilla 110-V, Valpara\'iso, Chile}
\affiliation[c]{Department of Particle Physics, Tel Aviv University, Tel Aviv 69978, Israel}
\affiliation[d]{Physics Department, Ben-Gurion University of the Negev, Beer Sheva 84105, Israel}

\abstract{Further developing ideas set forth in \cite{KLL}, we  discuss  QCD Reggeon Field Theory (RFT) and formulate restrictions imposed on its Hamiltonian by the  unitarity of underlying  QCD. We identify explicitly the QCD RFT Hilbert space, provide algebra of the basic degrees of freedom (Wilson lines and their duals) and the algorithm for calculating the scattering amplitudes. We formulate conditions imposed on the ``Fock states" of RFT by unitary nature of  QCD, and explain how these constraints appear as unitarity constraints on possible RFT hamiltonians that generate energy evolution of scattering amplitudes. We study the realization of these constraints in the dense-dilute limit of RFT where the appropriate Hamiltonian is the JIMWLK Hamiltonian $H_{JIMWLK}$. We find that the action $H_{JIMWLK}$ on the dilute projectile states is unitary, but acting on dense ``target" states it violates unitarity and generates states with negative probabilities through energy evolution. %In an attempt to address this problem we propose a family of RFT Hamiltonians, which, on the one hand reduce to  $H_{JIMWLK}$ in the dense-dilute limit, and on the other
}

\begin{document}
\maketitle

\pagestyle{empty}
\newpage

\mbox{}

\pagestyle{plain}

\setcounter{page}{1}
\author{}

\abstract{ }
\dedicated{}
\preprint{}
%\end{center}
%\bigskip

%\begin{abstract}

%\begin{document}
\section{Introduction.}
In this paper we continue the study of the s-channel unitarity of Reggeon Field Theory (RFT) of Quantum Chromodynamics (QCD). We further develop ideas put forward in \cite{KLL} on the restrictions that unitarity of QCD imposes on the Hamiltonian of RFT.

Reggeon Field Theory (RFT) is an effective theory for description of hadronic scattering in QCD at asymptotically high energies. The basic ideas of RFT go back to Gribov \cite{gribov},
 and have been developed over the years in the context of QCD \cite{BFKL,glr,MUPA,MUDI,LIREV,LipatovFT, bartels,BKP, mv,Salam, KOLE,BRN,braun,BK}.   In its modern form, the QCD RFT in a certain limit has been identified \cite{reggeon}  with the so called JIMWLK evolution equation \cite{jimwlk}, or Color Glass Condensate (CGC)\cite{cgc}. The relevant limit is when a perturbative dilute projectile scatters on a dense target.

 % Subsequently further relation between the CGC based approach and the RFT was explored. In particular recently we have shown that  one can generalize the JIMWLK Hamiltonian consistently in the regime where large Pomeron Loops are important \cite{ourbraun}. This regime includes the evolution of an initial dilute-dilute scattering to large rapidities, where at any given intermediate rapidity at most one of the evolved systems is dense. In this regime only Pomeron (Reggeon) splittings are important close to either one of the colliding objects, and one can write down a Hamiltonian, which encompasses both JIMWLK, and its dual (KLWMIJ\cite{klwmij}) evolution. The Hamiltonian in this regime contains the two triple Pomeron vertices, and (in the large $N_C$ limit) is the CGC equivalent of the Pomeron Lagrangian proposed by Braun \cite{braun} for description of the problem of scattering of large (but dilute) nuclei. So far, neither CGC nor RFT has been formulated in the most general case of scattering of two dense objects, although some work in this direction has been done\cite{KLduality,Balitsky05,SMITH,foam,GLV}.

 There are some significant differences between the original Gribov RFT framework and its QCD incarnation. The original reggeons in Gribov's RFT  are colorless, whereas the effective high energy degrees of freedom in QCD are colored  reggeized  gluons \cite{gluon} or Wilson lines.  It  must be possible "to integrate over the color" and  reformulate QCD RFT in terms of  color neutral  exchange amplitudes, such as BFKL Pomeron \cite{BFKL}. 
 % without violating any basic properties of QCD, particularly its unitarity.  
When written in terms of color singlet objects, the QCD RFT in addition to the Pomeron contains  higher
order colorless Reggeons, such as quadrupoles and higher multipoles which may play important role especially far from the dense-dilute limit. 

%Whether these higher Reggeons significantly affect high energy behavior of QCD amplitudes is not known at present. Finally, even if  the higher Reggeons can be discarded, it is not known whether the effective Pomeron Field Theory
%has a finite number of transition vertices. The large $N_C$ limit of high energy QCD is a convenient setup for the study of these questions. In this paper we stick to the large $N_c$ limit and in fact restrict ourselves even further by considering the dipole model approach\cite{MUDI}, in which the Pomeron is the only relevant degree of freedom at high energy.

  The CGC formalism provides a direct link between the fundamental theory (QCD) and the effective high energy description  (RFT). This prompted us to analyze in \cite{KLL}  some peculiar properties of RFT solutions found in the literature\cite{brauntarasov,motyka} from the point of view of the underlying QCD structure. 
  
The conclusions of \cite{KLL} were somewhat unexpected, but to our mind interesting and important. We found that the peculiar behavior of solutions of RFT found in \cite{brauntarasov,motyka}  is due to violation of the QCD s-channel unitarity in the implementations of RFT widely practiced in the community\cite{braun}. By considering the zero dimensional toy model of the RFT -type,  in \cite{KLL} we have explicitly showed how unitarity conditions are violated and also have found a way to modify the toy model RFT Hamiltonian such that  it remains consistent with all known limits, but the unitarity of the evolution is restored.

The full QCD RFT is of course significantly more complicated than the zero dimensional toy model. It was argued in \cite{KLL} that the unitarity is indeed violated also in this case for the Balitsky-Kovchegov (BK) \cite{BK} and Braun \cite{braun} Hamiltonians. However the explicit construction of the RFT algebra and the action of the appropriate Hamiltonians on RFT states were not provided, and thus the discussion lacked any detailed understanding of the origin of this violation.

 In the present paper we extend this analysis by providing an explicit and detailed formulation of the QCD RFT Hilbert space and operator algebra. As opposed to the toy model case \cite{KLL}, where we have used the large $N_c$ limit in which the Pomeron is the only relevant degree of freedom in RFT, here we deal with the full color structure.  Thus the focus of the current paper is the JIMWLK Hamiltonian $H_{JIMWLK}$ \cite{jimwlk,cgc} which generates the QCD evolution of scattering amplitudes in the dense-dilute case.% We also suggest a  self dual generalization of $H_{JIMWLK}$,  under dense-dilute duality transformation introduced in \cite{KLduality}. 
 %However at this point we are unable to analyze the unitarity properties of this extension.

 We analyze in detail the evolution of the projectile and target within this dense-dilute RFT framework. We show that while the evolution of the dilute object (projectile) is unitary, that of the dense object (target) is not. Moreover the violation of unitarity is qualitatively very similar to that in the toy model and derives directly from the assumption that every target gluon scatters at most via two gluon exchange. This approximation is perfectly adequate for the evolution of the scattering amplitude for as long as the projectile remains dilute, but breaks down once both scattering objects become dense. Note however, that the evolution of the scattering amplitude is distinct from the evolution of the state (see below). The early symptom of the impending breakdown of the evolution of the amplitude is indeed the unitarity violation in the evolution of the target state even while the scattering matrix evolution is still governed by the JIMWLK Hamiltonian.

The plan of this paper is as follows. In Section II we recap the unitarity problem and its solution in the zero dimensional toy model analyzed in \cite{KLL}. In Section III we discuss the general framework of QCD RFT in  2+1 (transverse+rapidity) dimensions. As mentioned above, we do not restrict ourselves to large $N_C$ limit and thus the basic degrees of freedom are the Wilson lines, or reggeized gluons. We define the algorithm of calculating the scattering amplitudes in this framework and define commutation relations between the projectile and target Wilson lines. 
In Section IV we explain in detail how the QCD s-channel unitarity appears in the context of RFT and what constraints it imposes on the evolution Hamiltonian. In Section V we analyze the JIMWLK limit in terms of its unitarity properties, and show that the (dilute) projectile evolution is unitary while the (dense) target evolution is not. We also explicitly demonstrate that this violation of unitarity arises due to restriction of the scattering amplitude of each individual target gluon to at most two gluon exchanges. Finally in Section VI we discuss our results and future prospects.
%In Section VI we discuss the set of self dual RFT Hamiltonians and show that they all reduce to the JIMWLK Hamiltonian in the dense-dilute limit. 

%We are motivated to consider this question by several works which explored numerical and analytic solutions of the Braun Pomeron Lagrangian\cite{brauntarasov,motyka}. 
%When the scattering amplitude is evolved within this framework to high enough rapidity, it exhibits paradoxical behavior which is very hard to understand from the point of view of the QCD ``black disk'' limit. 
%.

\section{The toy model - a recap.}
In this section we recap the main results of \cite{KLL} pertaining to a zero dimensional toy model. This explains our main idea in a simple setting and sets the stage for the discussion of QCD.
\subsection{The toy model RFT}

Consider the prototype of the Reggeon Field Theory defined in zero transverse dimensions. The only degree of freedom of this theory is the scalar Pomeron "field" $P$ and its dual $\bar P$, or alternatively "dipole fields" $d=1-P;\ \ \ \ \bar d=1-\bar P$. A useful way of thinking about these objects is that of a scattering amplitude of a target dipole on the projectile and the projectile dipole on the target respectively. In this toy model we are dealing only with the color singlet objects - "dipoles" which simplifies things considerably.

Mathematically the RFT is defined in terms of the following three elements:

1. The algebra of $P$ and $\bar P$;

2. The algorithm for calculating scattering amplitudes;

3. The energy evolution of the scattering amplitudes.

The zero dimensional toy model frequently used in the literature is defined by the following realization of these three elements: 

1. The commutation relations of $P$ and $\bar P$ are based on their perturbative identification:
\begin{equation}\label{alp}
[P,\bar P]=-\gamma; \ \ \ \ \ \ \gamma\sim O(\alpha_s^2)
\end{equation}
where $\gamma$ is the zero dimensional proxy for the dipole-dipole scattering amplitude. 

2. The $S$-matrix element for the scattering of $\bar n$ dipoles of the projectile on $m$ dipoles of the target  is calculated as  \cite{KLremark2} 
\beq\label{toyamp}
\langle m|\bar n\rangle=\int d\bar P\delta(\bar P)(1-P)^m(1-\bar P)^{\bar n}
\eeq
This can be conveniently represented in an alternative form
\begin{equation}\label{toyampF}
\langle m|\bar n\rangle=\langle  0|(1-P)^m(1-\bar P)^{\bar n}|  0\rangle
\eeq
where the left and right Pomeron  "Fock space vacua" are defined by
\begin{equation}
\langle 0|\bar P=P|0\rangle=0
\end{equation}
or
\begin{equation}
\langle 0|\bar d=\langle 0|;\ \ \ \ \ \ \ \ d|0\rangle=|0\rangle
\end{equation}
Clearly the algorithms eq.(\ref{toyamp}) and eq.(\ref{toyampF}) are equivalent given eq.(\ref{alp}).

Within this framework the state
\begin{equation}
\langle m|\equiv\langle  0|d^m
\end{equation}
has the meaning of a target state with $m$ dipoles, and similarly  $|\bar n\rangle$ is the projectile state with $\bar n$ dipoles.  
Commuting the factors of $P$ through the factors of $\bar P$ all the way to the right intuitively emulates the scattering of the projectile dipoles on the target dipoles. Once all the $P$ are on the right, they disappear when acting on $|0\rangle$ and the result gives the $S$-matrix element for the scattering of $\bar n$ dipoles of the projectile  on $m$ dipoles of the target.

3. The $S$ matrix element is evolved in rapidity according to
\beq
\langle m|\bar n\rangle_Y=\langle  0|(1-P)^me^{H(P,\bar P)Y}(1-\bar P)^{\bar n}|0\rangle
\eeq
The simplest choice of the Hamiltonian which is commonly used in the toy model is the zero dimensional analog of the Balitsky - Kovchegov (BK)  Hamiltonian \cite{BK} 
\beq\label{hbk}
H_{BK}=-\frac{1}{\gamma}\left[ P\bar P- P\bar P^2\right]
\eeq
which contains a triple pomeron vertex. This zero dimensional RFT has been intensely studied as a simplified model to understand the physics of QCD RFT \cite{ACJ,AAJ,JEN,ABMC,CLR,CIAF,KOZLE,SHXI,BIT,nestor,LEPRI}.

\subsection{The unitarity condition}
Given this physical meaning of the RFT "Fock states",  \cite{KLL} has formulated the unitarity condition on the evolution Hamiltonian $H$. 
\begin{equation}\label{unip}
\langle m|e^{HY}=\sum_ia_i(Y)\langle i|;\ \ \ \ \ \ \ 1\ge a_i\ge 0; \ \ \ \ \sum_ia_i=1
\end{equation}
The meaning of this condition is simple and straightforward: when  a state with $m$ target dipoles is evolved in rapidity, the action of the evolution operator must result in another normalized state with expansion coefficients that have the property of probabilities. Note that the coefficients $a_i$ have the meaning of probabilities rather than amplitudes. The reason is  that in the large $N_c$ limit there is no interference between states with different numbers of dipoles, and thus the S-matrix of a superposition of dipole states is given by the weighted sum of the S-matrices of individual states with the weights given by probabilities to find the particular state in the original superposition.

The same condition has to be satisfied on the projectile side, since the Hamiltonian can be though as acting either on the projectile, or on the target.
\begin{equation}\label{unit}
e^{HY}|\bar n\rangle=\sum_i\bar a_i(Y)|i\rangle;\ \ \ \ \ \ \ 1\ge\bar a_i\ge 0; \ \ \ \ \sum_i\bar a_i=1
\end{equation}

 Both equations, eq.(\ref{unip}) and eq. (\ref{unit}) must be satisfied for the same Hamiltonian $H$.
\subsection{Unitarity violation}

It is a straightforward matter to check whether the evolution generated by the Hamiltonian $H_{BK}$ is unitary. Applying the evolution operator (over infinitesimal interval $Y=\Delta$) on the projectile and target states we find
\begin{equation}\label{projevolv}
%\langle m|H=-m\langle m|+m\langle m+1|; \ \ \ \ 
e^{\Delta H}|\bar n\rangle\approx  (1-\Delta \bar n )|\bar n \rangle|+\Delta \bar n |\bar n+1\rangle
 \end{equation}
 \begin{equation}\label{tarevolv}
 %&&H|\bar n\rangle=\bar n|\bar n\rangle -\bar n[1+\gamma(\bar n-1)]|\bar n-1\rangle+\gamma\bar n(\bar n-1)|\bar n-2\rangle; \nonumber\\
  \langle m|e^{\Delta H}=(1+\Delta m)\langle m| -\Delta m[1+\gamma(m-1)]\langle m-1|+\Delta\gamma m(m-1)\langle m-2|
 \end{equation}
Eq.(\ref{projevolv})  conforms with the unitarity constraints, as all the coefficients are positive and sum up to one. However the action of the evolution operator on the target clearly violates unitarity. Although the sum of all the coefficients is equal to unity, one of the coefficients in eq.(\ref{tarevolv}) is negative, and one is greater than unity. Another worrying feature of eq.(\ref{tarevolv}) is that the evolution seems to decrease the number of dipoles in the state, while on physics grounds we expect this number to grow just like for the projectile.

\subsection{Unitarity regained.}
A suspect feature of the Hamiltonian $H_{BK}$is that it is not symmetric between the target and the projectile\cite{KLduality}. Physically 
this is because it is suited for a situation where the projectile is dilute while the target is dense (dense-dilute limit). Thus even though one does expect the "correct" Hamiltonian to be symmetric (or self dual), the self duality is violated in the dense-dilute limit. One can wonder whether this lack of self duality is the reason for the unitarity violation in eq.(\ref{tarevolv}). The analysis of \cite{KLL} showed that just restoring self duality is not sufficient. E.g. adding an additional triple Pomeron vertex with the Pomeron and conjugate Pomeron interchanged does not solve the problem, but in fact exacerbates it.

Nevertheless \cite{KLL} came up with a variant of the original toy model which restores unitarity, and also reproduces correct dense-dilute limit. This "Unitarized Toy Model" Hamiltonian turned out also to be self dual.

To achieve this several modifications were introduced. The first modification concerns the commutation relations of $P$ and $\bar P$, or equivalently $d$ and $\bar d$. The mechanics of the calculation of the scattering amplitude within the RFT suggests that commuting $d$ through $\bar d$ one should pick up the factor equal to the $S$-matrix of dipole-dipole scattering. The commutation relations are therefore modified as follows
\begin{equation} \label{ald}
d\bar d=e^{-\gamma}\bar d d\approx (1-\gamma)\bar d d
\end{equation}
The second approximate equality is due to the smallness of $\gamma$ and is not crucial but convenient.
Note that for small $P$ and $\bar P$ (parametrically $1>P\bar P>\gamma$) the algebra eq.(\ref{ald}) reduces to eq.(\ref{alp}).

The second modification introduced in \cite{KLL} is to replace the Hamiltonian $H_{BK}$ by
\begin{equation}
H_{UTM}=-\frac{1}{\gamma}\bar P P=-\frac{1}{\gamma}(1-\bar d)(1-d)
\end{equation}
As explained in \cite{KLL}, in the limit when the projectile is dilute and target is dense, the Hamiltonian $H_{UTM}$ becomes equivalent to $H_{BK}$ as far as the evolution of scattering amplitude is concerned. However, as it is easy to check $H_{UTM}$ generates a  unitary evolution of both, the projectile and the target wave functions. A simple calculation yields
\beq \label{utmu}\langle m|e^{\Delta H_{UTM}}=\left[1-\frac{\Delta}{\gamma}[1-(1-\gamma)^{m}]\right]\langle m|+\frac{\Delta}{\gamma}[1-(1-\gamma)^{m}]\langle m+1|
\eeq
This evolution is clearly unitary. Due to self duality, it is clear that the evolution of the projectile wave function is unitary as well. An attractive feature of this evolution is that the number of dipoles in the target grows with energy, rather than decreases as in eq.(\ref{tarevolv}). 
 Interestingly it also exhibits the saturation behavior very similar to the one that is expected from the real QCD evolution, namely at large $m$, the change in the wave function is independent of the number of dipoles $m$.

Thus the unitarized toy model setup solves several issues inherent in the BK limit. The evolution is explicitly self dual, it reduces to BK evolution for the s-matrix in the dense-dilute limit and is unitary.  This evolution also has basic properties that we expect on physical grounds - the number of dipoles always grows with rapidity, and at large number of dipoles the evolution saturates in the sense that the probability to produce an extra  dipole does not depend on the number of dipoles already present in the wave function.

\section{The Reggeon Field Theory: scattering amplitudes and field algebra.}
We now move on to consider the high energy limit of QCD.

The aim of this section is to define the "Hilbert space" of the Reggeon Field Theory and rules for calculating of scattering amplitudes. As explained in the Introduction, the route we take is through "translating" the Color Glass Condensate (CGC) formalizm to the RFT language.

\subsection{The $S$-matrix.}
We start with the basic formula for calculating scattering amplitudes in the CGC approach as discussed for example in 
\cite{yin}.
\beq
\label{scatm}
{\cal S}=\int d\rho d\alpha_T \delta(\rho)W_P[R]e^{i\int_z g^2\rho^a(\mathbf{z})\alpha_T^a(\mathbf{z})}\tilde W_T[\alpha_T]=\int d\rho \delta(\rho)W_P[R]W_T[S]
\eeq

Here $\rho^a(x)$ is the color charge density of the projectile, $\alpha^a_T(x)$ is the color field of the target, and
$R$ and $S$ are defined as
\beq \label{RS}
R_x=e^{T^a\frac{\delta}{\delta\rho^a(\mathbf{x})}}; \ \ \ \ \ \ \ \ \ \ S_x=e^{igT^a\alpha^a(\mathbf{x})}
\eeq
with the projectile color field $\alpha^a(x)$ determined by the projectile color charge density $\rho^a(x)$ via solution of the static Yang-Mills equations.
The operator $R$ is the "dual Wilson line''. An insertion of a factor $R$ in the amplitude eq.(\ref{scatm}) is equivalent to appearance of an extra eikonal scattering factor associated with an additional parton in the projectile. In this sense $R$ creates an additional parton in the projectile wave function. The Wilson line $S$ involves the projectile color field and has the meaning of the eikonal $s$-matrix of a {\it target} parton that scatters on the projectile.  
Here we have denoted the functional Fourier transform of $\tilde W_T[\alpha_T]$ by $W_T[S]$. In the following we will use the notation that stresses the similar role of $R$ and $S$ and the duality between the two
\begin{equation}
\bar U(\mathbf{x})\equiv R(\mathbf{x});\ \ \ \ \ \ U(\mathbf{x})\equiv S(\mathbf{x})
\end{equation}

The structure of the weight functions $W_P$ and $W_T$ is crucially important for the subsequent discussion of unitarity. This structure has been discussed in detail \cite{yin,yinyang}. The presence of a physical gluon at a transverse position $\mathbf{x}$ in the projectile wave function is associated with the factor $\bar U(\mathbf{x})$ in $W_P$. Thus for a wave function that contains a distribution of gluon configurations (numbers and positions), the projectile weight function has the general form
\beq\label{wp}
W_P=\sum_{n,\{a,b;\mathbf{x}\}}F^n(\{a,b;\mathbf{x}\})\prod_{i=1}^n[\bar U^{a_ib_i}(\mathbf{x}_i)]
\eeq

The physical meaning of the  functions $F^n$  has been discussed in \cite{reggeon}. To clarify it let us 
consider the eikonal scattering of  a QCD projectile state which is a superposition of  the QCD Fock space states with gluons in transverse positions $\mathbf{x}_1,...,\mathbf{x}_n$ with color indices $a_1,...,a_n$. The initial QCD projectile state  is thus
\begin{equation}\label{psii}
|\Psi_i\rangle=\sum_{n; \mathbf{x}_i;a_i}C_{a_1,a_2...a_n}|\mathbf{x}_1,a_1;...;\mathbf{x}_n,a_n\rangle
\end{equation}
while the final state after the scattering we take to be
\begin{equation}
|\Psi_f\rangle=\sum_{n; \mathbf{x}_i;b_i}C_{b_1,b_2...b_n}|\mathbf{x}_1,b_1;...;\mathbf{x}_n,b_n\rangle
\end{equation}
In the eikonal approximation this is the most general final state allowed since neither the number of gluons nor their transverse positions  change during the scattering.
The s-matrix element for this process is calculated using eq.(\ref{scatm}) with the projectile weight function in eq.(\ref{wp}) with 
\begin{equation}\label{fn}
F^n(\{a,b;\mathbf{x}\})=C_{a_1,a_2...a_n}(\mathbf{x}_1...\mathbf{x}_n)C^*_{b_1,b_2...b_n}(\mathbf{x}_1...\mathbf{x}_n)
\end{equation}

This in particular means that for $\{a_i\}=\{b_i\}$ the function $F^n(a,a,x)$ has the meaning of the probability density, and therefore has to be positive
\beq\label{prob}
F^n(\{a,a;\mathbf{x}\})\ge 0
\eeq
and normalized
\beq
\sum_{n,\{a\}}\int_{\{\mathbf{x}\}}F^n(\{a,a;\mathbf{x}\})=1; \
\eeq
Note that eq.(\ref{prob}) is valid for a fixed value of indexes $a_i$ - there is no summation over the indexes.
These properties of the functions $F^n$ are the basis of the unitarity conditions discussed below.

Similarly, a gluon in the target wave function carries a factor 
$U(\mathbf{y})$ in $W_T$, so
\beq\label{wt}
W_T=\sum_{n,\{c,d;\mathbf{y}\}}\bar F^n(\{c,d;\mathbf{x}\})\prod_{i=1}^n[ U^{c_i,d_i}(\mathbf{y}_i)]
\eeq
with the constraint
\beq
\bar F^n(\{b,b;\mathbf{y}\})\ge 0
\eeq
and normalization
\beq
\sum_{n,\{b\}}\int_{\{\mathbf{y}\}}\bar F^n(\{b,b;\mathbf{y}\})=1; \
\eeq

The calculation of the scattering amplitudes in RFT lends itself to a similar representation as in the toy model. Define the left and right Fock vacuum states by
\begin{equation}
\langle L|\bar U_{ab}=\delta_{ab}\langle L|; \ \ \ \ \ \ U_{ab}|R\rangle=\delta_{ab}|R\rangle
\end{equation}
Then the s-matrix element for scattering from the initial state $|\Psi_i\rangle=|\mathbf{x}_1,a_1;...;\mathbf{x}_N,a_N\rangle_T|\mathbf{y}_1, c_1;...;\mathbf{y}_{M}, c_{M}\rangle_P$ to the final state $|\Psi_f\rangle=|\mathbf{x}_1,b_1;...;\mathbf{x}_N,b_N\rangle_T|\mathbf{y}_1, d_1;...;\mathbf{y}_{M}, d_{M}\rangle_P$ is given by
\begin{equation}\label{sif}
S_{if}=\langle L| U^{a_1b_1}(\mathbf{x}_1) \ldots U^{a_Nb_N}(\mathbf{x}_N)\bar{U}^{c_1d_1}(\mathbf{y}_1) \ldots \bar{U}^{c_Md_M}(\mathbf{y}_M)|R\rangle
\end{equation}

\subsection{The algebra}

Considered as operators on the space of functionals $W$, the objects $U$ and $\bar U$ have nontrivial commutation relations. In principle those are directly calculable from the definitions eq.(\ref{RS}) but this is not a trivial calculation.  In the literature these commutation relations are usually approximated by those calculated in the dilute regime. In this regime, where any projectile gluon scatters only on a single target gluon (and vice versa) the commutator is given by the perturbative scattering amplitude \cite{reggeon}. This is the analog of the perturbative commutation relation eq.(\ref{alp}) in the toy model. Our goal is to determine the algebra of $U$ and $\bar U$ that goes beyond this perturbative expression.

To do this we use eq.(\ref{RS}). Since $\alpha$ is determined by $\rho$ through the solution of classical Yang Mills equation, in principle this can be done directly. There is one subtlety though related to the gauge invariance of QCD. The color charge density $\rho^a$ is not a gauge invariant object, and thus the algebra of $U$ and $\bar U$ in principle depends on the gauge chosen. This is not a problem in principle, as the commutator of $U$ and $\bar U$ must reproduce summation of perturbation theory diagrams, and we know that different sets of diagrams contribute in different gauges. Here we will choose the simplest possible setting. In particular it was shown in \cite{kovchegov} that in the Lorentz gauge the classical equation of motion for the color field is given by the first order perturbative expression
\begin{equation}\label{lorentz}
\alpha^a(\mathbf{x})=\int_{\mathbf{y}}\phi(\mathbf{x-y})\rho^a(\mathbf{y});\ \ \ \ \ \ \ \ \ \ \ \ \phi(\mathbf{x-y})=\frac{g}{2\pi}\ln \frac{|\mathbf{x-y}|}{L}
\end{equation}
The scale $L$ is arbitrary and does not enter calculations of any physical quantities.
Additionally the advantage of this gauge is that it is explicitly symmetric between the projectile and the target, and thus provides the simplest environment for realization of the duality transformation.

With this choice the Wilson line (reggeized gluon) operators become
\begin{equation}\label{algebra}
\bar U(\mathbf{x})=e^{T^a\frac{\delta}{\delta\rho^a(\mathbf{x})}}\ ; \ \ \ \ \ \ \ \ \ \  U(\mathbf{x})=e^{igT^a\int_y\phi(\mathbf{x-y})\rho^a(\mathbf{y})}
\end{equation}

These equations imply non-trivial commutation relations, which constitute the algebra of the RFT in analogy with eq.(\ref{ald}) in the toy model. Eq.(\ref{algebra}) is more complicated than in the zero dimensional case for two reasons: first,  because the basic fields carry color index, and second because the QCD interaction is nonlocal in the transverse space. Nevertheless they provide explicit realization of the algebra of the fundamental RFT fields in the RFT Hilbert space.

The commutation relation of the Wilson lines in (\ref{algebra}) has a simple diagrammatic interpretation. Consider for example the scattering of one gluon on one gluon. The scattering amplitude up to second order in $\alpha_s$ is given by 
\begin{equation}
\langle L| U^{ab}(\mathbf{x}) \bar{U}^{cd}(\mathbf{y}) |R\rangle =\delta^{ab}\delta^{cd}- ig\phi(\mathbf{x}-\mathbf{y}) T^i_{ab}T^i_{cd} +\left[ \frac{1}{2!}ig\phi(\mathbf{x}-\mathbf{y})\right]^2  (T^iT^j)_{ab}[(T^iT^j)_{cd} + (T^jT^i)_{cd}]+ \ldots
\end{equation}
This corresponds to the sum of  one and two gluon exchange diagrams in Fig. 1-a.
In fact it is easy to see by explicit calculation that higher order terms  organize themselves into all possible diagrams where the relative order of the vertices on the target gluon line is permuted in all possible ways. The $O(\alpha_s^3)$ contributions correspond to the three gluon exchange diagrams in Fig.1-b.

%%%%%%%%%%%%%%%%%%%%%%%%%%%%%%%%%%%%%%%%%%%%%%%%%%%
\begin{figure}[t]   
\centering   
\begin{tabular}{c}
  \includegraphics[width=9cm] {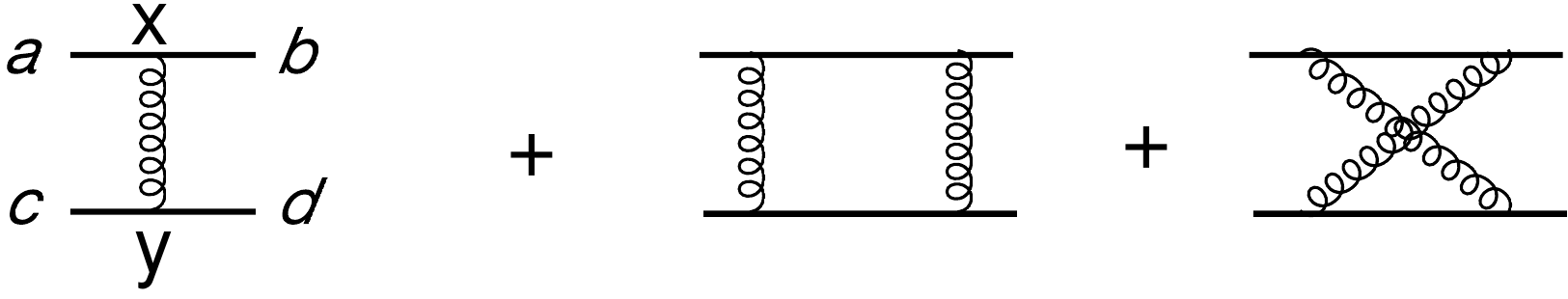} \\ 
  \fig{1}-a\\
  ~~\\
   \includegraphics[width=18cm] {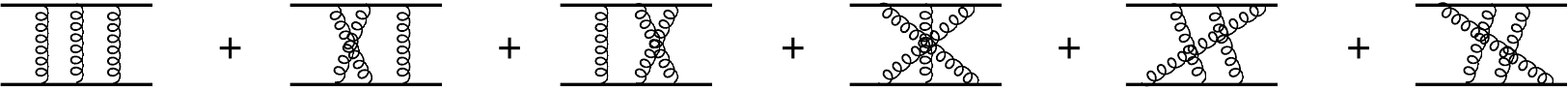}  \\
    \fig{1}-b
         \end{tabular}
\caption{The one, two (\fig{1}-a) and three(\fig{1}-b) gluon exchange contributions to the algebra.}
\label{1}
\end{figure}
%%%%%%%%%%%%%%%%%%%%%%%%%%%%%%%%%%%%%%%%%%%%%%%%%%

The one on two gluon scattering amplitude in the two gluon exchange approximation  is given by
\begin{eqnarray}
&&\langle L| U^{ab}(\mathbf{x}) \bar{U}^{c_1d_1}(\mathbf{y_1})\bar{U}^{c_2d_2}(\mathbf{y_2}) |R\rangle = \delta^{ab}\delta^{c_1d_1}\delta^{c_2d_2}-ig\left[\phi(\mathbf{x}-\mathbf{y_1})T^i_{c_1d_1}\delta_{c_2d_2}+ \phi(\mathbf{x}-\mathbf{y_2})T^i_{c_2d_2}\delta_{c_1d_1}\right]T^i_{ab} \nonumber\\
&&+\left[ \frac{1}{2!}ig\phi(\mathbf{x}-\mathbf{y_1})\right]^2  (T^iT^j)_{ab}[(T^iT^j)_{c_1d_1} + (T^jT^i)_{c_1d_1}]\delta_{c_2d_2}\nonumber\\
&&+ \left[ \frac{1}{2!}ig\phi(\mathbf{x}-\mathbf{y_2})\right]^2  (T^iT^j)_{ab}[(T^iT^j)_{c_2d_2} + (T^jT^i)_{c_2d_2}]\delta_{c_1d_1}\nonumber\\
&&-\frac{1}{2!}g^2\phi(\mathbf{x}-\mathbf{y_1})\phi(\mathbf{x}-\mathbf{y_2}) (T^iT^j)_{ab}[T^i_{c_1d_1}T^j_{c_2d_2} +T^i_{c_2d_2}T^j_{c_1d_1}]+\ldots 
\end{eqnarray}
Diagrammatically this is depicted in Fig.2. It is now clear what types of diagrams are encoded in the algebra of (\ref{algebra}).

With the algebra encoded in (\ref{algebra}) and the rule for calculating scattering amplitudes eq.(\ref{sif}) the framework of the QCD RFT is defined. To complete the RFT framework  one needs to specify the Hamiltonian  $H_{RFT}$ that generates the evolution of the scattering amplitude in energy. This Hamiltonian currently is known in the limit where the projectile is dilute, and the target is dense - the so called dense-dilute limit. We will define this Hamiltonian below. But before setting along this route  we will formalize the 
unitarity constraints on the energy evolution imposed by requiring that the energy evolution of the scattering amplitude is a manifestation of a unitary evolution of the QCD wave function of a hadronic state.

\begin{figure}[t]   \centering                                    
                                  \includegraphics[width=14cm] {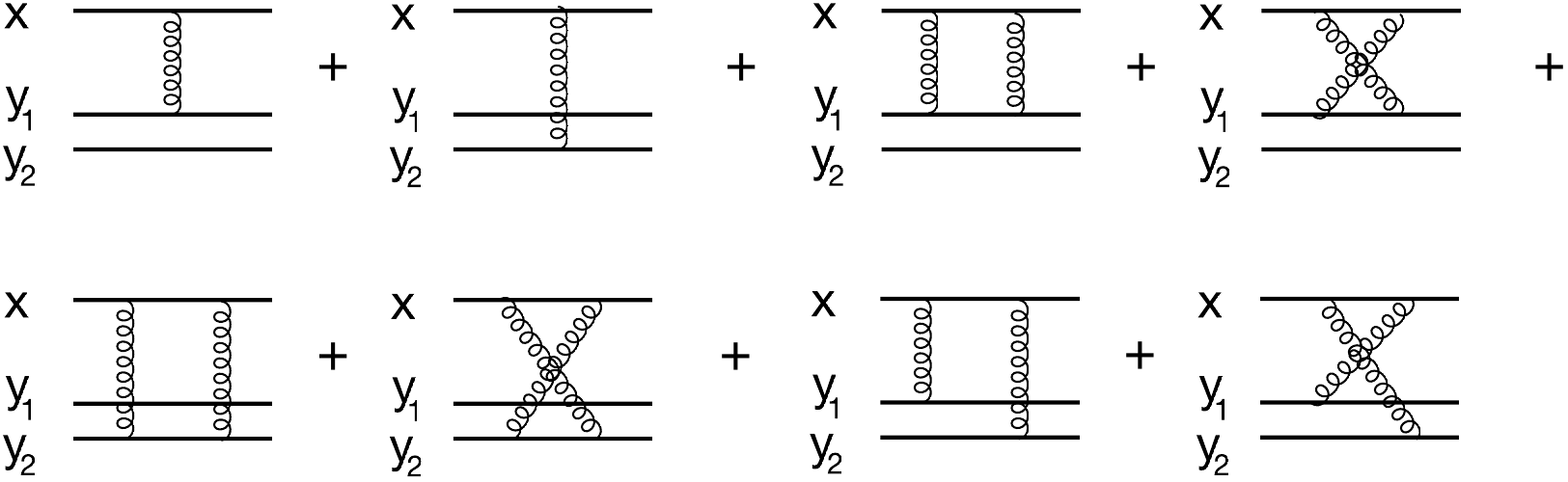}             
\caption{The one and two gluon exchange contributions to the algebra to one on two scattering.}
\label{2}
\end{figure}

\section{QCD unitarity and the RFT evolution}

%So far we have described the mechanics of calculation of scattering amplitudes within our framework of RFT. The next question to tackle is how the scattering amplitudes evolve with energy. 
In general the energy evolution is generated by the action of the RFT Hamlitonian $H_{RFT}[U,\bar U]$. The s-matrix element of eq.(\ref{sif}) evolved to rapidity $Y$ is given by
\begin{equation}\label{sev}
S_{if}(Y)=\langle L| U^{a_1b_1}(\mathbf{x}_1) \ldots U^{a_Nb_N}(\mathbf{x}_N)e^{YH_{RFT}[U,\bar U]}\bar{U}^{c_1d_1}(\mathbf{y}_1) \ldots \bar{U}^{c_Md_M}(\mathbf{y}_M)|R\rangle
\end{equation}
Although eq.(\ref{sev}) gives the evolution of the scattering amplitude, our discussion in the previous section allows us to reinterpret it in terms of the evolution of the wave function.

In particular let us consider only the target part of this expression. This can be interpreted  as  the evolution of the target RFT state
\begin{eqnarray}
&&\langle L| U^{a_1b_1}(\mathbf{x}_1) \ldots U^{a_Nb_N}(\mathbf{x}_N)\rightarrow \langle L| U^{a_1b_1}(\mathbf{x}_1) \ldots U^{a_Nb_N}(\mathbf{x}_N)e^{YH_{RFT}}\\
=&&\sum_{n,\{\bar a,\bar b;\mathbf{\bar x}\}}F_N^n(Y,\{a,b,\mathbf{x}; \ \bar a,\bar b,\mathbf{\bar x}\})\langle L|\prod_{i=1}^n[U^{\bar a_i\bar b_i}(\mathbf{\bar x}_i)]\nonumber
\end{eqnarray}
Here to calculate the right hand side of the equality one has to commute all the operators $\bar U$ that are present in the evolution operator $e^{YH_{RFT}}$ all the way to the left next to $\langle L|$, at which point they disappear and the resulting expression by definition can be written as a superposition of states with different numbers of the projectile gluons. The coefficients in this  superposition, the functions $F_N^n$ depend on the initial state being evolved and the evolution parameter $Y$. The summation in principle goes over all possible gluon numbers $n$ from zero to infinity as well as all possible transverse positions and color indexes of these gluons.

This evolution of the RFT state must be inherited from the unitary evolution to higher energy of the QCD wave function with $N$ gluons. The QCD wave function that results from the evolution must of course satisfy all the properties of a wave function of a normalized state in the Hilbert space, i.e. of the form eq.(\ref{psii}). Therefore the coefficients
$F_N^n$ must also be of the form eq.(\ref{fn}).

We are therefore lead to the following conditions imposed by the unitarity of QCD evolution:
\begin{equation}\label{pu}
1>F_N^n(Y,\{a,a,\mathbf{x}; \ \bar a,\bar a,\mathbf{\bar x}\})>0; \ \ \ \ \sum_n\sum_{\bar a}\int \{d\mathbf{\bar x}\} F_N^n(Y,\{a,a,\mathbf{x}; \ \bar a,\bar a,\mathbf{\bar x}\})=1
\end{equation}
Note that there is no summation over $a_i$ or $\bar a_i$ in the first of these equations, while in the second only indexes $\bar a_i$  are summed over. The transverse coordinates $\{\mathbf{x}_i\}$ are fixed in both equations as well.

We will refer to this as the target unitarity condition.

This discussion can be repeated {\it verbatum} for the projectile. In eq.(\ref{sev}) we can act with the evolution operator on the projectile state and generate the evolved projectile RFT state
\begin{eqnarray}
&&\bar{U}^{c_1d_1}(\mathbf{y}_1) \ldots \bar{U}^{c_Md_M}(\mathbf{y}_M)|R\rangle
\rightarrow e^{YH_{RFT}[U,\bar U]}\bar{U}^{c_1d_1}(\mathbf{y}_1) \ldots \bar{U}^{c_Md_M}(\mathbf{y}_M)|R\rangle\\
=&&\sum_{m,\{\bar c,\bar d;\mathbf{\bar y}\}}G_M^m(Y,\{c,d,\mathbf{y}; \ \bar c,\bar d,\mathbf{\bar y}\})\prod_{i=1}^m[\bar U^{\bar c_i\bar d_i}(\mathbf{\bar y}_i)]|R\rangle\nonumber
\end{eqnarray}

The coefficients $G_M^m$ are subject to the projectile unitarity condition
\begin{equation}\label{tu}
1>G_M^m(Y,\{c,c,\mathbf{y}; \ \bar c,\bar c,\mathbf{\bar y}\})>0; \ \ \ \ \sum_m\sum_{\bar c}\int \{d\mathbf{\bar y}\} G_M^m(Y,\{c,c,\mathbf{y}; \ \bar c,\bar c,\mathbf{\bar y}\})=1
\end{equation}
Both equations eq.(\ref{pu}) and eq.(\ref{tu}) have to be satisfied simultaneously with the same Hamiltonian $H_{RFT}$.

Our procedure to check whether the unitarity conditions are satisfied will be based on the following steps.

1. Given $H_{RFT}$, act with it on the projectile wave function dispensing of all the operators $ U$ by commuting them all the way to the right.

2. Represent the resulting expression as expansion in powers of $\bar U(\mathbf{y})$, and identify the expansion coefficients $G_N^n$.

3. Verify that the diagonal coefficients $G_N^n(c,c,\mathbf{y}; \bar c,\bar c,\mathbf{\bar y};)$ satisfy eq.(\ref{tu}).

4. Repeat the procedure for the target.

\section{Unitarity violation in JIMWLK evolution}
\subsection{The JIMWLK limit}
To complete our setup of RFT we need to specify the Hamiltonian of the evolution $H_{RFT}$. This Hamiltonian is known in the limit when the projectile is dilute and the target is dense. In this limit $H_{RFT}$ is given by the JIMWLK Hamiltonian.
\begin{equation}\label{hjimwlk}
H_{JIMWLK}=\frac{\alpha_s}{2\pi^2}\int_{\mathbf{x,y,z}}\frac{(\mathbf{x-z})\cdot(\mathbf{y-z})}{ \mathbf{(x-z)^2(y-z)^2}}\left[2\mathcal{J}_L^a(\mathbf{x})\mathcal{J}_R^b(\mathbf{y}) \bar U^{ab}(\mathbf{z})-\mathcal{J}_L^a(\mathbf{x})\mathcal{J}_L^a(\mathbf{y})-\mathcal{J}_R^a(\mathbf{x})\mathcal{J}_R^a(\mathbf{y})\right]
\end{equation}
The right and left rotation operators are defined as \cite{ddd}
%To construct the RFT Hamiltonian, we first introduce the following operators
\begin{equation}\label{j}
\begin{split}
&\mathcal{J}_L^a(\mathbf{x})  = \rho^b(\mathbf{x}) \left[\frac{1}{2}T^e\frac{\delta}{\delta \rho^e(\mathbf{x})} \left(\coth{\left[\frac{1}{2}T^e\frac{\delta}{\delta \rho^e(\mathbf{x})}\right]}-1\right)\right]^{ba}\, ,\\
&\mathcal{J}_R^a(\mathbf{x})  = \rho^b(\mathbf{x}) \left[\frac{1}{2}T^e\frac{\delta}{\delta \rho^e(\mathbf{x})} \left(\coth{\left[\frac{1}{2}T^e\frac{\delta}{\delta \rho^e(\mathbf{x})}\right]}+1\right)\right]^{ba}\, ,\\
\end{split}
\end{equation}

The function on the right hand side as usual should be understood as a power series expansions. For a single variable $t$ we define
\begin{equation}
\begin{split}
&M_L(t)\equiv\frac{t}{2}\left(\coth{\frac{t}{2}} -1\right) = \frac{t}{e^t-1} =\sum_{m=0}^{\infty} \frac{B^-_m t^m}{m!} = \sum_{m=0}^{\infty} C^-_m t^m\\
&M_R(t)\equiv\frac{t}{2}\left(\coth{\frac{t}{2}} +1\right) = \frac{t}{1-e^{-t}} =\sum_{m=0}^{\infty} \frac{B^+_m t^m}{m!} =  \sum_{m=0}^{\infty} C^+_m t^m\\
\end{split}
\end{equation}
Here $B_m^-$ and $B_m^+$ are Bernoulli numbers. They have the properties that $B_{2n}^-=B_{2n}^+$ for all even integers $2n$ while $B_{2n+1}^-=B_{2n+1}^+=0$ for all odd integers $2n+1$ except $B_1^- = -\frac{1}{2} = -B_1^+$. Also the relations $M_L(t) = M_R(t) e^{-t}$ and $M_R(t) = M_L(t) e^{t}$ can be readily verified.

The operators $\mathcal{J}_L^a(\mathbf{x}), \mathcal{J}^a_R(\mathbf{x})$ act as left rotation and right rotation on the Wilson line $\bar{U}^{mn}(\mathbf{x})$, 
\begin{equation}\label{eq:LR_rot_barV}
\begin{split}
&[\mathcal{J}_L^a(\mathbf{x}), \bar{U}^{mn}(\mathbf{y})] = - (T^a \bar{U}(\mathbf{y}))^{mn}\delta(\mathbf{x}-\mathbf{y})\, ,\\
&[\mathcal{J}_R^a(\mathbf{x}), \bar{U}^{mn}(\mathbf{y})] = - ( \bar{U}(\mathbf{y})T^a)^{mn}\delta(\mathbf{x}-\mathbf{y})\,. \\
\end{split}
\end{equation}
and satisfy the $SU(N)\times SU(N)$ commutation relations:
\begin{equation}
\begin{split}
&[\mathcal{J}_L^a(\mathbf{x}), \mathcal{J}_L^b(\mathbf{y})] = if^{abc} \mathcal{J}^c_L(\mathbf{x})\delta(\mathbf{x}-\mathbf{y})\,, \\
&[\mathcal{J}_R^a(\mathbf{x}), \mathcal{J}_R^b(\mathbf{y})] = -if^{abc} \mathcal{J}^c_R(\mathbf{x})\delta(\mathbf{x}-\mathbf{y})\, .\\
\end{split}
\end{equation}
\begin{equation}
[\mathcal{J}_L^a(\mathbf{x}), \mathcal{J}_R^a(\mathbf{y})] =0\, .
\end{equation}

%For future use let us also define analogous rotation operators for $U$.
%\begin{equation}
%\begin{split}
%&\mathcal{I}_L^a(\mathbf{x})  = \frac{-i}{g}\frac{\delta}{\delta \alpha^b(\mathbf{x})}  \left[\frac{1}{2}T^eig\alpha^e(\mathbf{x})\left(\coth{\left[\frac{1}{2}T^eig\alpha^e(\mathbf{x})\right]}-1\right)\right]^{ba}\, ,\\
%&\mathcal{I}_R^a(\mathbf{x})  = \frac{-i}{g}\frac{\delta}{\delta \alpha^b(\mathbf{x})}  \left[\frac{1}{2}T^eig\alpha^e(\mathbf{x})\left(\coth{\left[\frac{1}{2}T^eig\alpha^e(\mathbf{x})\right]}+1\right)\right]^{ba}\, ,\\
%&\mathcal{I}_L^a(\mathbf{x})  = \frac{-i}{g^2}\partial^2 \frac{\delta}{\delta \rho^b(\mathbf{x})}  \left[\frac{1}{2}T^eig^2\frac{1}{\partial^2} \rho^e(\mathbf{x})\left(\coth{\left[\frac{1}{2}T^eig^2\frac{1}{\partial^2} \rho^e(\mathbf{x})\right]}-1\right)\right]^{ba}\, ,\\
%&\mathcal{I}_R^a(\mathbf{x})  = \frac{-i}{g^2}\partial^2 \frac{\delta}{\delta \rho^b(\mathbf{x})}  \left[\frac{1}{2}T^eig^2\frac{1}{\partial^2} \rho^e(\mathbf{x})\left(\coth{\left[\frac{1}{2}T^eig^2\frac{1}{\partial^2} \rho^e(\mathbf{x})\right]}+1\right)\right]^{ba}\, .\\
%\end{split}
%\end{equation}
%with $\alpha^a(\mathbf{x})=g\left[\frac{1}{\partial^2}\rho^a\right](\mathbf{x})=\int_{\mathbf{y}}\phi(\mathbf{x-y})\rho^a(\mathbf{y})$.
%These satisfy
%\begin{equation}
%\begin{split}
%&[U^{mn}(\mathbf{y}), \mathcal{I}^a_L(\mathbf{x})] = - (T^a U(\mathbf{y}))^{mn}\delta(\mathbf{x}-\mathbf{y})\, ,\\
%&[U^{mn}(\mathbf{y}), \mathcal{I}^a_R(\mathbf{x})] = - ( U(\mathbf{y})T^a)^{mn}\delta(\mathbf{x}-\mathbf{y})\, .\\
%\end{split}
%\end{equation}

The Hamiltonian eq.(\ref{hjimwlk}) has been derived in several approaches directly from the fundamental QCD theory in the dense-dilute regime. In the present context the dense-dilute regime means that one of the scattering objects, say projectile contains a small number of gluons, $\bar m\sim 1$, while the other contains a parametrically large number $n\sim 1/\alpha_s^2$.

%Note that $\mathcal{I}$ and $\mathcal{J}$  are related by the (canonical) dense-dilute duality transformation 

As first discussed in \cite{KLduality} the full RFT Hamiltonian should be invariant under the dense-dilute duality transformation
\begin{equation}
\rho^a(\mathbf{x})\leftrightarrow -\frac{i}{g}\frac{\delta}{\delta \alpha^a(\mathbf{x})}
\end{equation}
Physically this transformation corresponds to interchanging the projectile and the target. It should be the symmetry of the RFT Hamiltonian, as it is just the matter of choice which one of colliding objects one calls projectile, and which one target \cite{KLduality}. The Hamltonian eq.(\ref{hjimwlk}) is clearly not symmetric under this transformation. This is not surprising since it is designed to describe a clearly asymmetric situation where the target and projectile qualitatively differ from each other. The situation is similar to that in the zero dimensional toy model, where we saw that the lack of self duality meant that the action of the Hamiltonian on the projectile and target wave functions is very different.

%The $\mathcal{J}_L^a(\mathbf{x}), \mathcal{J}^a_R(\mathbf{x}), \mathcal{I}_L^a(\mathbf{x}), \mathcal{I}_R^a(\mathbf{x})$ are constructed as functionals of $\rho^a(\mathbf{x})$ and $\delta/\delta \rho^a(\mathbf{x})$, which are fundamental degrees of freedom of the theory. In $\mathcal{I}_L^a(\mathbf{x}), \mathcal{I}_R^a(\mathbf{x})$, 
%\begin{equation}
%ig^2\frac{1}{\partial^2} \rho^e(\mathbf{x}) = ig^2 \int d^2\mathbf{y} \phi(\mathbf{x}-\mathbf{y}) \rho^e(\mathbf{y}) \, .
%\end{equation}
%Also note the commutation relations
%\begin{equation}
%\begin{split}
%&\left[ \frac{\delta}{\delta \rho^a(\mathbf{x})}, \rho^b(\mathbf{y})\right] = \delta^{ab}\delta(\mathbf{x}-\mathbf{y})\, ,\\
%&\left[ig^2\frac{1}{\partial^2} \rho^a(\mathbf{x}), \frac{-i}{g^2}\partial^2 \frac{\delta}{\delta \rho^b(\mathbf{y})} \right] = -\delta^{ab}\delta(\mathbf{x}-\mathbf{y}) \, .\\
%\end{split}
%\end{equation}

%Likewise, $\mathcal{I}_L^a(\mathbf{x}), \mathcal{I}_R^a(\mathbf{x})$ serves as left and right rotations of $U^{mn}(\mathbf{x})$, 

In this section we study the action of the JIWMLK Hamiltonian on the projectile and target wave functions. We will see that the situation again mirrors that of a toy model, i.e. the action of $H_{JIMWLK}$ on the dilute wave function is unitary but on the dense one is not.

Let us first consider the action on a dilute projectile.

\subsection{Dilute projectile evolution is unitary.}
Since the JIMWLK Hamiltonian contains products of only two rotation generators it can act at most at two factors of $\bar U$ in the projectile wave function. Thus without loss of generality we consider the projectile that only contains two such factors. A little algebra gives
\begin{equation}\label{ondilute}
\begin{split}
&H_{JIMWLK} \bar{U}^{p_1q_1}(\mathbf{y}_1)\bar{U}^{p_2q_2}(\mathbf{y}_2)\\
=&\frac{1}{2\pi}\int_{\mathbf{x}}N_c[i\partial_x\phi(\mathbf{x}-\mathbf{y}_1)]^2 \bar{U}^{p_1q_1}(\mathbf{y}_1) \bar{U}^{p_2q_2}(\mathbf{y}_2)+N_c [i\partial_x\phi(\mathbf{x}-\mathbf{y}_2)]^2 \bar{U}^{p_1q_1}(\mathbf{y}_1)\bar{U}^{p_2q_2}(\mathbf{y}_2) \\
&+2[i\partial_x\phi(\mathbf{x}-\mathbf{y}_1)][i\partial_x\phi(\mathbf{x}-\mathbf{y}_2)][T^e\bar{U}(\mathbf{y}_1)]^{p_1q_1}[T^e\bar{U}(\mathbf{y}_2)]^{p_2q_2}\\
&+N_c[i\partial_x\phi(\mathbf{x}-\mathbf{y}_1)]^2 \bar{U}^{p_1q_1}(\mathbf{y}_1) \bar{U}^{p_2q_2}(\mathbf{y}_2)+ N_c[i\partial_x\phi(\mathbf{x}-\mathbf{y}_2)]^2  \bar{U}^{p_1q_1}(\mathbf{y}_1)\bar{U}^{p_2q_2}(\mathbf{y}_2)\\
&+2[i\partial_x\phi(\mathbf{x}-\mathbf{y}_1)][i\partial_x\phi(\mathbf{x}-\mathbf{y}_2)][\bar{U}(\mathbf{y}_1)T^e]^{p_1q_1}[\bar{U}(\mathbf{y}_2)T^e]^{p_2q_2}\\
&-2\bar{U}^{ed}(\mathbf{x}) [i\partial_x\phi(\mathbf{x}-\mathbf{y}_1)]^2[T^e\bar{U}(\mathbf{y}_1)T^d]^{p_1q_1}\bar{U}^{p_2q_2}(\mathbf{y}_2)-2\bar{U}^{ed}(\mathbf{x}) [i\partial_x\phi(\mathbf{x}-\mathbf{y}_2)]^2[T^e\bar{U}(\mathbf{y}_2)T^d]^{p_2q_2}\bar{U}^{p_1q_1}(\mathbf{y}_1)\\
&-2\bar{U}^{ed}(\mathbf{x})[i\partial_x\phi(\mathbf{x}-\mathbf{y}_1)][i\partial_x\phi(\mathbf{x}-\mathbf{y}_2)] \left([T^e\bar{U}(\mathbf{y}_1)]^{p_1q_1}[\bar{U}(\mathbf{y}_2)T^d]^{p_2q_2}+ [T^e\bar{U}(\mathbf{y}_2)]^{p_2q_2}[\bar{U}(\mathbf{y}_1)T^d]^{p_1q_1}\right)
\end{split}
\end{equation}
First consider terms that contain two factors of $\bar{U}$. Those arise from the virtual term in $H_{JIMWLK}$. To extract probabilities for these states, we set $\bar{U}\rightarrow I$ and focus on diagonal elements $p_1=q_1, p_2=q_2$.  Note that single gluon exchange pieces like $[T^e\bar{U}(\mathbf{y}_1)]^{p_1q_1}$ and $[T^e\bar{U}(\mathbf{y}_2)]^{p_2q_2}$ do not contribute to forward scattering amplitudes and thus vanish once we set $\bar U$ equal to unit matrix.  
Recall that we are interested in the wave function after  evolution over a small rapidity interval  $\Delta$, i.e. we should consider $\exp[\Delta H_{JIMWLK}]\approx 1+\Delta H_{JIMWLK}$. The probability to find the two gluon state in the evolved wave function is therefore (this probability does not depend on the value of color indexes $p_1$ and $p_2$)
\begin{equation}
\mathcal{P}_{(2)}(\mathbf{y}_1, \mathbf{y}_2)=1-\frac{\Delta}{\pi}  N_c\int_{\mathbf{x}} \left([\partial_x\phi(\mathbf{x}-\mathbf{y}_1)]^2+[\partial_x\phi(\mathbf{x}-\mathbf{y}_2)]^2\right)\, .
\end{equation}
which for small $\Delta$ is positive and smaller than unity.

Terms that contain three factors of $\bar{U}$ represent three gluon states in the evolved wave function. Separating these terms, setting $\bar{U}$ to unity and focusing on $p_1=q_1, p_2=q_2$, one obtains the probability for three gluon state with color indices $p_1,p_2,e$ 
\begin{equation}
\mathcal{P}_{(3)}(\mathbf{x},e; \mathbf{y}_1,p_1, \mathbf{y}_2,p_2) = \frac{\Delta}{\pi}  \left([\partial_x\phi(\mathbf{x}-\mathbf{y}_1)]^2\sum_qf^{ep_1q}f^{e,p_1q}+[\partial_x\phi(\mathbf{x}-\mathbf{y}_2)]^2\sum_qf^{ep_2q}f^{e,p_2q}\right)
\end{equation}
The emitted gluon associated with the factor $\bar{U}^{ed}(\mathbf{x})$ can be at an arbitrary transverse position $\mathbf{x}$. Different values of $\mathbf{x}$ correspond to orthogonal components of the wave function containing gluons at different transverse positions. The probabilities, $\mathcal{P}_{(3)}(\mathbf{x}; \mathbf{y}_1, \mathbf{y}_2)$ have to be positive for arbitrary ${\mathbf x}$ and arbitrary values of color indexes, and indeed they  are. When integrating over  $\mathbf{x}$ and summing over color index $e$ we get  $\mathcal{P}_{(2)}+\int_{\mathbf{x}}\sum_e\mathcal{P}_{(3)}=1$  thus conserving the total probability.  As expected the evolution of a dilute projectile preserves unitarity.

As a popular example, consider the dilute projectile to be a dipole composed of two gluons. 
\begin{equation}
d(\mathbf{y}_1, \mathbf{y}_2) = \frac{1}{N_c^2-1}\mathrm{Tr}[\bar{U}(\mathbf{y}_1)\bar{U}^{\dagger}(\mathbf{y}_2)]\, .
\end{equation}
Note that in the adjoint representation $[T^j\bar{U}(\mathbf{y}_2)]^{p_1q_1} = -[\bar{U}^{\dagger}(\mathbf{y}_2)T^j]^{q_1p_1}$.

The evolution in this case is simply the right hand side of the BK equation:
\begin{equation}
\delta d(\mathbf{y}_1, \mathbf{y}_2) =\frac{\Delta}{\pi}\left[\int_{\mathbf{x}} \mathcal{K}(\mathbf{x},\mathbf{y}_1, \mathbf{y}_2) (\mathrm{Tr}\left[\bar{U}(\mathbf{y}_1)\bar{U}^{\dagger}(\mathbf{y}_2)\right] - \mathrm{Tr}\left[T^j\bar{U}(\mathbf{y}_1)\bar{U}^{\dagger}(\mathbf{x}) T^j \bar{U}(\mathbf{x}) \bar{U}^{\dagger}(\mathbf{y}_2)\right])\right]
\end{equation}
with
\begin{equation}
\begin{split}
\mathcal{K}(\mathbf{x},\mathbf{y}_1, \mathbf{y}_2) &= \left( [i\partial_x\phi(\mathbf{x}-\mathbf{y}_1)]^2 +  [i\partial_x\phi(\mathbf{x}-\mathbf{y}_2)]^2-2[i\partial_x\phi(\mathbf{x}-\mathbf{y}_1)][i\partial_x\phi(\mathbf{x}-\mathbf{y}_2)]\right) \\
&= -\frac{g^2}{(2\pi)^2}\frac{(\mathbf{y}_1-\mathbf{y}_2)^2}{(\mathbf{x}-\mathbf{y}_1)^2(\mathbf{x}-\mathbf{y}_2)^2}
\end{split}
\end{equation}
Clearly probabilities of the new state components (extra gluon at $\mathbf{x}$ with an arbitrary color index) are positive and add up to unity.

\subsection{Unitarity Violation for  Dense Target }
We now study evolution of the dense target wave function within JIMWLK approximation.
While the action of $H_{JIMWLK}$ on the projectile is straightforward, understanding how it acts on the dense target presents a significant challenge. For example, $\mathcal{J}_L$ and $\mathcal{J}_R$ act on $\bar U$ as a simple left and right rotation. On the other hand to act with either of them on $U$ one has to realize the infinite number of derivatives in eq.(\ref{j}). Nevertheless, as we will see below we can make some headway using the fact that while scattering on a dilute projectile each target gluon can exchange no more than two gluons.

 To facilitate the calculation we first reexpress the JIMWLK Hamiltonian in a different form using integration by parts. %given that $\phi(\mathbf{x}-\mathbf{y})$ approaches zero fast enought as $\mathbf{x}\rightarrow 0$. 
We start with eq.(\ref{hjimwlk}) which we write as  
\begin{equation}\label{byparts}
\begin{split}
H_{JIMWLK}   
 =& \frac{1}{2\pi}\int_{\mathbf{x},\mathbf{y},\mathbf{z}}\partial^2_{\mathbf{x}}[i\phi(\mathbf{x}-\mathbf{y}) i\phi(\mathbf{x}-\mathbf{z})]\left[-2\bar{U}^{ed}(\mathbf{x})\mathcal{J}_L^e(\mathbf{y})\mathcal{J}_R^d(\mathbf{z}) + \mathcal{J}^e_L(\mathbf{y})\mathcal{J}^e_L(\mathbf{z}) + \mathcal{J}^e_R(\mathbf{y})\mathcal{J}^e_R(\mathbf{z}) \right]\nonumber\\
\end{split}
\end{equation}
We now integrate by parts over $\mathbf{x}$. Upon integration by parts the terms not involving $\bar{U}(\mathbf{x})$ vanish, and we obtain
 %, the first line of eq. \eqref{eq:twoEps_JIMWLK}, one obtains
 %\begin{equation}\label{eq:secondExp_JIMWLK}
 %\begin{split}
%H_{JIMWLK} &  = \frac{1}{4}\int_{\mathbf{x},\mathbf{y},\mathbf{z}}\partial^2_{\mathbf{x}}[i\phi(\mathbf{x}-\mathbf{y}) i\phi(\mathbf{x}-\mathbf{z})]\left[-2\bar{U}^{ed}(\mathbf{x})\mathcal{J}_L^e(\mathbf{y})\mathcal{J}_R^d(\mathbf{z}) \right]\\
%&=\frac{1}{2}\int_{\mathbf{x},\mathbf{y},\mathbf{z}}[i\partial_{\mathbf{x}}\phi(\mathbf{x}-\mathbf{y}) i\partial_{\mathbf{x}}\phi(\mathbf{x}-\mathbf{z})]\left[-2\bar{U}^{ed}(\mathbf{x})\mathcal{J}_L^e(\mathbf{y})\mathcal{J}_R^d(\mathbf{z}) \right] \\
%&\qquad+ \frac{1}{2} \int_{\mathbf{y},\mathbf{z}} \phi(\mathbf{y}-\mathbf{z}) \left[ \mathcal{J}_L^e(\mathbf{y})\mathcal{J}_L^e(\mathbf{z}) + \mathcal{J}_R^e(\mathbf{y})\mathcal{J}_R^e(\mathbf{z})\right]\\
%\end{split}
%\end{equation}
%The second equality is related to the second line of eq. \eqref{eq:twoEps_JIMWLK} via integration by parts. 
 
%Carrying out integration by parts in the first line of eq. \eqref{eq:secondExp_JIMWLK}, one obtains a third expression for JIMWLK Hamiltonian
\begin{equation}\label{jimwlkf}
H_{JIMWLK}   = \frac{1}{2\pi}\int_{\mathbf{x},\mathbf{y},\mathbf{z}}\left[-2\partial_{\mathbf{x}}^2\bar{U}^{ed}(\mathbf{x})\right][i\phi(\mathbf{x}-\mathbf{y})\mathcal{J}_L^e(\mathbf{y})][ i\phi(\mathbf{x}-\mathbf{z})\mathcal{J}_R^d(\mathbf{z}) ]
\end{equation}
This form of the JIWMLK Hamitonian will be our starting point for analyzing its action on a dense object\footnote{
Note that in order to be able to perform integration by parts in eq.(\ref{byparts}) we need to assume that $\phi(\mathbf{x}-\mathbf{y})
\xrightarrow{\mathbf{x}\rightarrow\infty}0$. This implies that the scale $L$ in eq.(\ref{lorentz}) has to be set equal to the linear size of the system. Although this is perfectly acceptable, we point out that for amplitudes which involve the {\bf projectile} in globally gauge invariant initial and final state this does not matter. Changing the scale $L$ amounts to shifting $\phi$ by a constant in eq.(\ref{jimwlkf}) which  generates terms proportional to $\int_{\mathbf{y}}\mathcal{J}_L^e(\mathbf{y})$ and $\int_{\mathbf{y}}\mathcal{J}_R^e(\mathbf{y})$. These  vanish for a globally color singlet projectile. The projectile color singlet condition poses no restrictions on possible target states, and thus we will continue to deal with all possible states in the Fock space basis of the target.}.

Our goal now is to act with the Hamiltonian in eq.(\ref{jimwlkf}) on a string of matrices $U$ representing the target state. First, we recall that in the dense-dilute limit each factor $U$ can only exchange two gluons with whatever object it scatters on. Practically this means that we should expand every factor of $U$ in the wave function to second order in $\rho$. The truncated form of $U$ is
\begin{equation}\label{uweak}
U(x)=1+igT^a\alpha^a(\mathbf{x})-\frac{g^2}{2}T^aT^b\alpha^a(\mathbf{x})\alpha^b(\mathbf{x})
\end{equation}
This although seemingly a trivial matter has an interesting effect.  After experiencing two gluon exchanges any given target gluon cannot scatter anymore and thus disappears from the wave function. As we will see this has a resut that $H_{JIMWLK}$ when acting on the target actually {\bf annihilates} gluons present in the dense wave function - the effect opposite to its action on the dilute wave function, where it {\bf creates} new gluons see eq.(\ref{ondilute}). This state of affairs is very similar to the zero dimensional toy model we have discussed in Section II.

So how do we in practice evaluate the action of the Hamiltonian on the string of $U$'s? The general idea is the following. Using eqs.(\ref{algebra},\ref{j}) we express $H_{JIMWLK}$ as a function of $\delta/\delta\rho^a$ and $\rho^a$. We then act with the derivatives on the string of $U$'s in the straightforward manner. This gives us a function of $\rho^a(\mathbf{x})$, or equivalently $\alpha^a(x)$. We then express these factors of $\alpha^a(\mathbf{x})$ in terms of $U$ essentially inverting eq.(\ref{uweak}). The result then is a function of $U$'s only, which now can be written as a series in powers of $U$ and the unitarity condition can be analyzed.

The procedure outlined above in principle allows one to calculate probabilities of all the Fock components of the evolved wave function. However in practice this involves very lengthy algebra. We will not pursue this in full generality, but will rather only calculate explicitly "probabilities" of two Fock space components: we start with the target state that has $N$ gluons, allow it to evolve over very short evolution interval $\Delta$, and calculate the probability of the evolved state to have $N$ and $N-1$ gluons. Even this calculation is rather technical, but we feel that it is a "necessary evil" and present the main steps in this section.
At the end of the day we will find that the probability for $N-1$ gluon state is negative, while that of the original $N$ gluon state is not changed by the evolution. This unambiguously establishes the violation of unitarity.

We start with the following simple observation. Any factor of $U$ in the target wave function after scattering on the Hamiltonian (i.e. being acted upon by the Hamiltonian) will have to be set to unit matrix for the purpose of extracting probabilities. This means that it has to either not scatter on $H_{JIMWLK}$ at all, or scatter via a singlet two gluon exchange. If it exchanges only one gluon with the Hamiltonian, its contribution will vanish upon setting $U=1$. Mathematically this is just the statement that for $U$ given by eq.(\ref{uweak}) one has
\begin{equation}
\frac{\delta}{\delta \rho^a(\mathbf{x})}U^{bb}(\mathbf{y})|_{\alpha=0}=1
\end{equation}
Thus the derivative $\delta/\delta\rho^a$ in $H_{JIMWLK}$ will always act in pairs on factors of $U$, and each such pair will annihilate one $U$. 
After utilizing all derivatives $\delta/\delta\rho^a$ in $H_{JIMWLK}$ we will be left with two factors of $\alpha^a(\mathbf{x})$ contained in $\mathcal{J}_L^e$ and $\mathcal{J}_R^e$. These two factors have to be combined into $U$ according to eq.(\ref{uweak}). Note that the two factors have to be combined into {\it the same} $U$. If each $\alpha(\mathbf{x})$ represents a single gluon exchange of a separate $U$, this contribution will vanish after setting $U=1$ for the  same reason as discussed above. This means that for the calculation of the probabilities only those terms in $H_{JIMWLK}$ matter which have the two factors of $\alpha(\mathbf{x})$ at the same transverse coordinate.  Such terms can be extracted by rewriting 
\begin{equation}
\begin{split}
\int_{\mathbf{y}}i\phi(\mathbf{x}-\mathbf{y})\mathcal{J}_L^e(\mathbf{y})& = \int_{\mathbf{y}}i\phi(\mathbf{x}-\mathbf{y}) \rho^p(\mathbf{y}) M_L^{pe}\left[ T^a\delta/\delta\rho^a(\mathbf{y})\right]\\
&=\frac{1}{g}\int_{\mathbf{y}}i\phi(\mathbf{x}-\mathbf{y}) \partial_{\mathbf{y}}^2\alpha^p(\mathbf{y}) M_L^{pe}\left[ T^a\delta/\delta\rho^a(\mathbf{y})\right]\\
&=\frac{1}{g}\int_{\mathbf{y}}i\partial_{\mathbf{y}}^2\phi(\mathbf{x}-\mathbf{y}) \alpha^p(\mathbf{y}) M_L^{pe}\left[ T^a\delta/\delta\rho^a(\mathbf{y})\right]+\frac{1}{g}\int_{\mathbf{y}}i\phi(\mathbf{x}-\mathbf{y}) \alpha^p(\mathbf{y}) \partial_{\mathbf{y}}^2M_L^{pe}\left[ T^a\delta/\delta\rho^a(\mathbf{y})\right] \\
&\qquad +\frac{2}{g}\int_{\mathbf{y}}i\partial_{\mathbf{y}}\phi(\mathbf{x}-\mathbf{y}) \alpha^p(\mathbf{y})\partial_{\mathbf{y}} M_L^{pe}\left[ T^a\delta/\delta\rho^a(\mathbf{y})\right].\\
\end{split}
\end{equation}
Using $\partial^2\phi(\mathbf{x-y})=g\delta^2(\mathbf{x-y})$ we see that only in the first term  the transverse coordinate of $\alpha$ is the same as that of $\mathcal{J}_L$ . Therefore only this term will contribute to the calculation of probabilities. Thus {\bf for the purpose of calculation of probabilities only} we use
\begin{equation}
\int_{\mathbf{y}}i\phi(\mathbf{x}-\mathbf{y})\mathcal{J}_L^e(\mathbf{y})\simeq i \alpha^p(\mathbf{x}) M_L^{pe}\left[ T^a\delta/\delta\rho^a(\mathbf{x})\right]
\end{equation}
%In the second equality, we have used the relations $\partial_{\mathbf{x}}^2 \alpha^a(\mathbf{x}) = \rho^a(\mathbf{x})$, which is the same as $ \alpha^a(\mathbf{x}) = \int_{\mathbf{y}}\phi(\mathbf{x}-\mathbf{y}) \rho^a(\mathbf{y})$.  We performed integration by parts in obtaining the third equality and ignored the boundary terms. In the last line, we only keep terms that involve $\alpha^a(\mathbf{x})$ at transverse position $\mathbf{x}$. Eventually, we will relate $\alpha^p(\mathbf{x})$ to \textit{adjoint} Wilson lines which represent the scattering amplitude of gluons.  Mathematically speaking, one can relate $\alpha^p(\mathbf{x})$ to any representations of $SU(N_c)$. For example, one single $\alpha^p(\mathbf{x})$ could be approximated by the Wilson line in the fundamental representation through $
%\alpha^p(\mathbf{x}) \simeq -i2\mathrm{Tr}(t^p V(\mathbf{x})) = i2\mathrm{Tr}(t^p V^{\dagger}(\mathbf{x})) $.  It can also be related to the Wilson lines in the adjoint representation by $\alpha^p(\mathbf{x}) = -i \mathrm{Tr} (T^p U(\mathbf{x}))/N_c$. Two $\alpha^{p}(\mathbf{x})$ at different transverse positions like $\alpha^{p}(\mathbf{y}) \alpha^q(\mathbf{z})$ are related to two different gluonic Wilson lines by $\mathrm{Tr} (T^p U(\mathbf{x}))\mathrm{Tr}(T^q U(\mathbf{x}))$ within the two gluon exchange approximation. These terms however will not contribute to forward scatterings.  On the other hand, if they have the same transverse positions, they are related to one gluonic Wilson line. 
Following similar argumentation, we also obtain
\begin{equation}
\begin{split}
\int_{\mathbf{z}}i\phi(\mathbf{x}-\mathbf{z})\mathcal{J}_R^e(\mathbf{z})& = \int_{\mathbf{z}}i\phi(\mathbf{x}-\mathbf{z}) \rho^q(\mathbf{z}) M_R^{qe}\left[ T^a\delta/\delta\rho^a({\mathbf{z}})\right]\\
&\simeq i \alpha^q(\mathbf{x}) M_R^{qe}\left[ T^a\delta/\delta\rho^a(\mathbf{x})\right].\\
\end{split}
\end{equation}

We thus consider the simplified version of the JIMWLK Hamiltonian that is equivalent to $H_{JIMWLK}$ as far as  extracting probabilities of the evolved states\footnote{In writing \eqref{eq:fundamental_to_adjoint}, we have ignored the terms arising from the action of $\mathcal{J}^d_R$ on $\mathcal{J}_L^e$ etc. within the Hamiltonian, as these terms contain a single power of $\alpha$ and thus again do not contribute to probabilities.}
\begin{equation}\label{eq:fundamental_to_adjoint}
\tilde{H}_{JIMWLK}   =\frac{1}{2\pi} \int_{\mathbf{x}}\left[2\partial_{\mathbf{x}}^2\bar{U}^{ed}(\mathbf{x})\right]M_L^{pe}\left[ T^a\delta/\delta\rho^a(\mathbf{x})\right]M_R^{qd}\left[ T^a\delta/\delta\rho^a(\mathbf{x})\right]\alpha^p(\mathbf{x})\alpha^q(\mathbf{x})
\end{equation}

In the approximation eq.(\ref{uweak}) the relation between the product $\alpha^p(\mathbf{x})\alpha^q(\mathbf{x}) $ and the one gluon Wilson line is
\begin{equation}\label{eq:2alpha_to_U}
\begin{split}
g^2\alpha^p(\mathbf{x})\alpha^q(\mathbf{x}) 
%&\simeq  2 \mathrm{Tr}(t^p V(\mathbf{x}))\mathrm{Tr}(t^q V^{\dagger}(\mathbf{x})) + (p\leftrightarrow q)\\
%&=2t^p_{ji}t^q_{lk} V_{ij}(\mathbf{x})V_{kl}^{\dagger}(\mathbf{x}) + (p\leftrightarrow q)\\
&=\frac{2}{N_c} \delta^{pq} + 4U^{ab}(\mathbf{x}) \left[\mathrm{Tr}(t^at^qt^bt^p) + \mathrm{Tr}(t^at^pt^bt^q)\right]\\
\end{split}
\end{equation}
where $t^a$ are generators in the fundamental representation.
%From the second line to the third line, we used the relation between fundamental Wilson lines and the adjoint Wilson lines in eq. \eqref{eq:fundamental_to_adjoint}.  
This can be explicitly verified
%eq. \eqref{eq:2alpha_to_U} is also explicitly checked 
by expanding $U^{ab}(\mathbf{x}) = e^{ig\alpha^m(\mathbf{x})T^m}$ to second order in $\alpha^m(\mathbf{x})$, and is done in Appendix A. 

%From eq. \eqref{eq:2alpha_to_U}, it looks like the $2/N_c\delta^{pq}$ does not change of the number of gluons in t dense projectile while the $4U^{ab}(\mathbf{x}) \left[\mathrm{Tr}(t^at^qt^bt^p) + \mathrm{Tr}(t^at^pt^bt^q)\right]$ {\bf ??????}

We are now ready to calculate probabilities to find a fixed number of gluons in the evolved  dense state
% when the JIMWLK Hamiltonian acts on dense projectiles $U^{a_1b_1}(\mathbf{x}_1)U^{a_2b_2}(\mathbf{x}_2)\ldots U^{a_Nb_N}(\mathbf{x}_N)$ containing $N$ gluons. 
 by considering  the expression
\begin{equation}\label{hu}
U^{a_1b_1}(\mathbf{x}_1)U^{a_2b_2}(\mathbf{x}_2)\ldots U^{a_Nb_N}(\mathbf{x}_N)H_{JIMWLK} .% \bar{U}^{c_1d_1}(\mathbf{y}_1)\bar{U}^{c_2d_2}(\mathbf{y}_2)\, .
\end{equation}

The Hamiltonian $\tilde H_{JIMLWK}$ contains infinite series in derivatives $\delta/\delta\rho$, which act in pairs on different factors of  $U$. The result of such an action is annihilating the Wilson line on which the two derivatives act
%Each $U^{a_lb_l}(\mathbf{x})$ is allowed to exchange two and only two gluons so that two $\delta/\delta\rho^a_{\mathbf{x}}$ kill one $U^{a_lb_l}(\mathbf{x})$ resulting in
\begin{equation}
\begin{split}
\frac{\delta}{\delta \rho^{c_1}_{\mathbf{x}}} \frac{\delta}{\delta \rho^{c_2}_{\mathbf{x}}} U^{a_lb_l}(\mathbf{x}_l)&= g^2\int_{\mathbf{y}_1, \mathbf{y}_2} \phi(\mathbf{x}-\mathbf{y}_1)\frac{\delta}{\delta \alpha^{c_1}_{\mathbf{y}_1}} \phi(\mathbf{x}-\mathbf{y}_2)\frac{\delta}{\delta \alpha^{c_2}_{\mathbf{y}_1}} \left( -\frac{1}{2!} (T^{a_1}T^{a_2}) \alpha^{a_1}_{\mathbf{x}_l} \alpha^{a_2}_{\mathbf{x}_l}\right)\\
&=  -g^2\frac{1}{2!} (T^{c_1}T^{c_2}+ T^{c_2}T^{c_1})_{a_lb_l}\phi(\mathbf{x}-\mathbf{x}_l)\phi(\mathbf{x}-\mathbf{x}_l)
\end{split}
\end{equation}
This is the consequence of the two gluon exchange approximation eq.(\ref{uweak})  inherent in the JIMWLK limit.
Since the expansion of $\tilde H_{JIMWLK}$ in the derivatives starts at the order $(\delta/\delta\rho)^2$, at least one gluon in the unevolved wave function is annihilated.
On the other hand the only "new" gluons are created by the factor $\alpha\alpha$, which according to eq.\eqref{eq:2alpha_to_U} creates at most one factor of $U$, or one additional gluon.
It is thus clear that the JIMWLK Hamiltonian does not increase the net number of gluons in the wave function of a dense target. Instead it generates Fock states with at most the same number of gluons as in the unevolved state.  This stands in stark contrast to its action on a dilute projectile such as $\bar{U}^{c_1d_1}(\mathbf{y}_1)\bar{U}^{c_2d_2}(\mathbf{y}_2)$, where it produces states with higher number of gluons.

%At most keeps the net number of gluons unchanged. The general situations are to decrease any number of gluons with the corresponding probabilities.  
Explicitly we have
\begin{equation}\label{eq:the_expansions_ML_MR_Ub}
\begin{split}
&\bar{U}^{ed}(\mathbf{x}) =\left[e^{T^a\frac{\delta}{\delta \rho^a_{\mathbf{x}}}} \right]^{ed}\\
&M_L^{pe}\left[ T^a\delta/\delta\rho^a(\mathbf{x})\right]  = 1 -\frac{1}{2} T^{c_1} \frac{\delta}{\delta \rho^{c_1}_{\mathbf{x}}} + \frac{1}{12} (T^{c_1}T^{c_2})\frac{\delta}{\delta \rho^{c_1}_{\mathbf{x}}} \frac{\delta}{\delta \rho^{c_2}_{\mathbf{x}}} - \frac{1}{720} (T^{c_1}T^{c_2}T^{c_3}T^{c_4})\frac{\delta}{\delta \rho^{c_1}_{\mathbf{x}}} \frac{\delta}{\delta \rho^{c_2}_{\mathbf{x}}}\frac{\delta}{\delta \rho^{c_3}_{\mathbf{x}}}\frac{\delta}{\delta \rho^{c_4}_{\mathbf{x}}}+\ldots\\
&M_R^{qd}\left[ T^a\delta/\delta\rho^a(\mathbf{x})\right]  = 1 +\frac{1}{2} T^{c_1} \frac{\delta}{\delta \rho^{c_1}_{\mathbf{x}}} + \frac{1}{12} (T^{c_1}T^{c_2})\frac{\delta}{\delta \rho^{c_1}_{\mathbf{x}}} \frac{\delta}{\delta \rho^{c_2}_{\mathbf{x}}} - \frac{1}{720} (T^{c_1}T^{c_2}T^{c_3}T^{c_4})\frac{\delta}{\delta \rho^{c_1}_{\mathbf{x}}} \frac{\delta}{\delta \rho^{c_2}_{\mathbf{x}}}\frac{\delta}{\delta \rho^{c_3}_{\mathbf{x}}}\frac{\delta}{\delta \rho^{c_4}_{\mathbf{x}}}+\ldots\\
\end{split}
\end{equation}
The terms, in which one of the derivatives $\delta/\delta\rho^a_{\mathbf{x}}$ comes from the expansion of $\bar{U}^{ed}$, contain one or two spatial derivatives:
\begin{equation}
\left(\partial^2_{\mathbf{x}}\frac{\delta}{\delta \rho^{c_1}_{\mathbf{x}}} \right)\frac{\delta}{\delta \rho^{c_2}_{\mathbf{x}}} U^{a_lb_l}(\mathbf{x}_l)=  -\frac{1}{2!} (T^{c_1}T^{c_2}+ T^{c_2}T^{c_1})_{a_lb_l}g^3\delta(\mathbf{x}-\mathbf{x}_l)\phi(\mathbf{x}-\mathbf{x}_l),
\end{equation}

\begin{equation}
\left(\partial_{\mathbf{x}}\frac{\delta}{\delta \rho^{c_1}_{\mathbf{x}}} \right)\frac{\delta}{\delta \rho^{c_2}_{\mathbf{x}}} U^{a_lb_l}(\mathbf{x}_l)=  -\frac{1}{2!} (T^{c_1}T^{c_2}+ T^{c_2}T^{c_1})_{a_lb_l}g^2\mathcal{B}_i(\mathbf{x}-\mathbf{x}_l)\phi(\mathbf{x}-\mathbf{x}_l),
\end{equation}

\begin{equation}
\left(\partial_{\mathbf{x}}\frac{\delta}{\delta \rho^{c_1}_{\mathbf{x}}} \right)\left(\partial_{\mathbf{x}}\frac{\delta}{\delta \rho^{c_2}_{\mathbf{x}}}\right) U^{a_lb_l}(\mathbf{x}_l)=  -\frac{1}{2!} (T^{c_1}T^{c_2}+ T^{c_2}T^{c_1})_{a_lb_l}g^2\mathcal{B}_i(\mathbf{x}-\mathbf{x}_l)\mathcal{B}_i(\mathbf{x}-\mathbf{x}_l)
\end{equation}
where $\mathcal{B}_i(\mathbf{x}-\mathbf{x}_l)=\frac{\partial}{\partial \mathbf{x}_i}\phi(\mathbf{x}-\mathbf{x}_l)$.

Only terms with even number of $\delta/\delta\rho^a_{\mathbf{x}}$ are kept in the $\tilde{H}_{JIMWLK}$ when combining different terms from the expansions of $\bar{U}^{ed}, M_L^{pe}$ and $M_R^{qd}$. 

\subsubsection{The $N$ Gluon Component of the Evolved State}
We now return to eq.(\ref{hu}) and calculate the probability to find an
 $N$ gluon state after the original $N$ gluon state has been evolved by a (infinitesimally) small rapidity interval $\Delta$.  To  do this we need to keep only those terms in $\tilde H_{JIMWLK}$ which do not change the number of gluons. Using eqs. \eqref{eq:fundamental_to_adjoint}, \eqref{eq:2alpha_to_U}, \eqref{eq:the_expansions_ML_MR_Ub}, these terms are
\begin{equation}
\begin{split}
\tilde{H}^{(N)}_{JIMWLK}=&\frac{1}{2\pi}\frac{1}{g^2}\int_{\mathbf{x}}\Bigg\{\partial_{\mathbf{x}}^2\left(T^{c_1}\frac{\delta}{\delta\rho^{c_1}_{\mathbf{x}}}\right)^{ed}\left[\delta^{pe}\left( \frac{1}{2}T^{c_2}\frac{\delta}{\delta\rho^{c_2}_{\mathbf{x}}}\right)^{qd} +\left(- \frac{1}{2}T^{c_2}\frac{\delta}{\delta\rho^{c_2}_{\mathbf{x}}}\right)^{pe}\delta^{qd} \right]\\
&+\partial_{\mathbf{x}}^2\left(\frac{1}{2}(T^{c_1}T^{c_2})\frac{\delta}{\delta\rho^{c_1}_{\mathbf{x}}}\frac{\delta}{\delta\rho^{c_2}_{\mathbf{x}}}\right)^{ed}\delta^{pe}\delta^{qd}\Bigg\}4U^{ab}(\mathbf{x}) \left[\mathrm{Tr}(t^at^qt^bt^p) + \mathrm{Tr}(t^at^pt^bt^q)\right]\\
=&\frac{1}{2\pi}\frac{1}{g^2}\int_{\mathbf{x}} -\frac{1}{2} (T^{c_1}T^{c_2} + T^{c_2}T^{c_1})_{pq} \left(\partial_{\mathbf{x}}^2\frac{\delta}{\delta\rho^{c_1}_{\mathbf{x}}}\right) \frac{\delta}{\delta\rho^{c_2}_{\mathbf{x}}} + \frac{1}{2}(T^{c_1}T^{c_2})_{pq} \left[\left(\partial^2_{\mathbf{x}}\frac{\delta}{\delta\rho^{c_1}_{\mathbf{x}}}\right)\frac{\delta}{\delta\rho^{c_2}_{\mathbf{x}}} + \frac{\delta}{\delta\rho^{c_1}_{\mathbf{x}}}\left(\partial^2\frac{\delta}{\delta\rho^{c_2}_{\mathbf{x}}}\right)\right]\\
&+ (T^{c_1}T^{c_2})_{pq}\partial_{\mathbf{x}}\frac{\delta}{\delta\rho^{c_1}_{\mathbf{x}}}\partial_{\mathbf{x}}\frac{\delta}{\delta\rho^{c_2}_{\mathbf{x}}}4U^{ab}(\mathbf{x}) \left[\mathrm{Tr}(t^at^qt^bt^p) + \mathrm{Tr}(t^at^pt^bt^q)\right]\\
=&\frac{1}{2\pi}\frac{1}{g^2}\int_{\mathbf{x}}(T^{c_1}T^{c_2})_{pq}\partial_{\mathbf{x}}\frac{\delta}{\delta\rho^{c_1}_{\mathbf{x}}}\partial_{\mathbf{x}}\frac{\delta}{\delta\rho^{c_2}_{\mathbf{x}}}4U^{ab}(\mathbf{x}) \left[\mathrm{Tr}(t^at^qt^bt^p) + \mathrm{Tr}(t^at^pt^bt^q)\right]\\
\end{split}
\end{equation}
Applying this to the $N$ gluon state we find
\begin{equation}
\begin{split}
&U^{a_1b_1}(\mathbf{x}_1)U^{a_2b_2}(\mathbf{x}_2)\ldots U^{a_Nb_N}(\mathbf{x}_N)\tilde{H}^{(N)}_{JIMWLK}\\
=&\sum_{l=1}^{N} U^{a_1b_1}(\mathbf{x}_1)\ldots U^{a_{l-1}b_{l-1}}(\mathbf{x}_{l-1})U^{a_{l+1}b_{l+1}}(\mathbf{x}_{l+1})\ldots U^{a_Nb_N}(\mathbf{x}_N) \frac{1}{2\pi}\int_{\mathbf{x}} (T^{c_1}T^{c_2})_{pq} \\
&\times \left[-\frac{1}{2!} (T^{c_1}T^{c_2}+ T^{c_2}T^{c_1})_{a_lb_l}\mathcal{B}_i(\mathbf{x}-\mathbf{x}_l)\mathcal{B}_i(\mathbf{x}-\mathbf{x}_l)\right]4U^{ab}(\mathbf{x}) \left[\mathrm{Tr}(t^at^qt^bt^p) + \mathrm{Tr}(t^at^pt^bt^q)\right]\\
\end{split}
\end{equation}
To extract probabilities, we set all the $U^{mn} \rightarrow \delta^{mn}$ and fix the indexes $m$. This gives the probability density of evolving of a Fock state $|\mathbf{x}_1,a_1;...\mathbf{x}_N,a_N\rangle$ into a state in which the original gluon at the transverse position $\mathbf{x}_l$ with color $a_l$ has been annihilated but at the same time a gluon at transverse position $\mathbf{x}$ with color $a$ is created. The probability density has the form
\begin{eqnarray}
&&\mathcal{P}_N(\{\mathbf{x}_i,a_i\}; \mathbf{x}_1,a_1...\mathbf{x}_{l-1},a_{l-1};\mathbf{x}_{l+1},a_{l+1}...\mathbf{x}_N,a_N; \mathbf{x},a)\\
=&&\frac{\Delta}{2\pi}(T^{c_1}T^{c_2})_{pq}  \left[-\frac{1}{2!} (T^{c_1}T^{c_2}+ T^{c_2}T^{c_1})_{a_la_l}\mathcal{B}_i(\mathbf{x}-\mathbf{x}_l)\mathcal{B}_i(\mathbf{x}-\mathbf{x}_l)\right]4 \left[\mathrm{Tr}(t^at^qt^at^p) + \mathrm{Tr}(t^at^pt^at^q)\right]\nonumber
\end{eqnarray}
for  $\mathbf{x}\ne\mathbf{x}_l$. Note that there is no summation over repeated color indices in the above expression. 
Performing the color algebra we have (see Appendix A) 
\begin{equation}\label{color}
\begin{split}
&-\frac{1}{2}(T^{c_1}T^{c_2})_{pq}  (T^{c_1}T^{c_2}+T^{c_2}T^{c_1})_{a_la_l} 4 \left[\mathrm{Tr}(t^at^qt^at^p) + \mathrm{Tr}(t^at^pt^at^q)\right]\\
=&-2 \left[\mathrm{Tr}(T^qT^{a_l}T^{a_l}T^p) +\mathrm{Tr}(T^pT^{a_l}T^{a_l}T^q)\right] \left[\mathrm{Tr}(t^at^qt^at^p) + \mathrm{Tr}(t^at^pt^at^q)\right]\\
=&-4 \left(-\frac{1}{2N_c} \delta^{aa}\delta^{a_la_l}+\frac{1}{2N_c} \delta_{aa_l}\delta_{aa_l} + \frac{1}{4} [(D^{a_l}D^{a_l})_{aa} -(T^{a_l}T^{a_l})_{aa}]\right)\\
\end{split}
\end{equation}
Here $D^{a}_{bc} = d_{abc}$ is the totally symmetric structure constant of $SU(N)$ algebra. 

The probability $\mathcal{P}_N$  has to be positive for any $\mathbf{x}$ and any value of indexes $a_l$ and $a$. To make things easier to understand we sum this expression over the index $a$.  This sum should also be positive if it is the sum of positive numbers. We then have
\begin{equation}\label{eq:U_becomes_delta}
4U^{ab}(\mathbf{x}) \left[\mathrm{Tr}(t^at^qt^bt^p) + \mathrm{Tr}(t^at^pt^bt^q)\right] \rightarrow -\frac{2}{N_c}\delta^{pq}
\end{equation}
and
\begin{eqnarray}
&&\sum_a\mathcal{P}_N(\{\mathbf{x}_i,a_i\}; \mathbf{x}_1,a_1...\mathbf{x}_{l-1},a_{l-1};\mathbf{x}_{l+1},a_{l+1}...\mathbf{x}_N,a_N; \mathbf{x},a)
 =  \frac{\Delta}{\pi}N_c  \mathcal{B}_i(\mathbf{x}-\mathbf{x}_l)\mathcal{B}_i(\mathbf{x}-\mathbf{x}_l)
\end{eqnarray}
This expression is positive for any $\mathbf{x}$ and $\mathbf{x}_l$ as the probability should be, and thus does not contradict unitarity.

However consider now the situation where $\mathbf{x}$ is equal to the transverse coordinate of one of the existing gluons and the color index $a=a_l$. The corresponding quantity $\mathcal{P}_N$ is then not a probability by itself, but the change in the probability to find the original configuration after the evolution. Unitarity requires this correction to be negative, as it has to cancel the contribution to the total probability due to all other states generated by the evolution.
However setting $a_l=a$ in eq.(\ref{color}) we find that this correction vanishes since $d_{aac}=0$ and $f_{aac}=0$.
We thus find that the correction to the original Fock state probability vanishes, which unambiguously means that unitary is violated.

To see that this violation indeed is reflected in negative probabilities, we now perform the calculation for a state with $N-1$ gluons.

%As long as the transverse coordinate $\mathbf{x}$ is distinct from the transverse coordinate of one of the existing gluons $\mathbf{x}_l$that for the added gluon $U^{ab}(\mathbf{x})$, different transverse positions $\mathbf{x}$ represent different gluon states. 

\subsubsection{The $N-1$ Gluon Component of the Evolved State}
There are two routes to obtain a state with $N-1$ gluons as a result of the evolution: one can either kill one gluon or kill two gluons and create one new gluon - the extra factor $U$ in the Hamiltonian. The former possibility lends itself to the same analysis as above. If the gluon at transverse position $\mathbf{x}_l$ is killed, one obtains the $N-1$ gluon state $|\mathbf{x}_1,a_1\ldots \mathbf{x}_{l-1},a_{l-1};\mathbf{x}_{l+1},a_{l+1}\ldots \mathbf{x}_N,a_N\rangle$ with the probability
\begin{equation}\label{pro0}
\mathcal{P}^0_{N-1}(\{\mathbf{x}_i,a_i\}; \mathbf{x}_1,a_1...\mathbf{x}_{l-1},a_{l-1};\mathbf{x}_{l+1},a_{l+1}...\mathbf{x}_N,a_N)
 = -2N_c  \frac{\Delta}{2\pi}\int_{\mathbf{x}} \mathcal{B}_i(\mathbf{x}-\mathbf{x}_l)\mathcal{B}_i(\mathbf{x}-\mathbf{x}_l)
\end{equation}
This expression follows directly from the result of the preceding subsection since the only difference in the calculation  is that this contribution comes from the first term rather than second term in eq.\eqref{eq:2alpha_to_U}. Using eq.\eqref{eq:U_becomes_delta} we see that the two coefficients are equal in magnitude and opposite in sign. 

Note that this contribution to probability is negative. 
This  is not the complete result yet. We need to consider also the second route of generating the $N-1$ gluon state.
This latter possibility is more complicated. We present the calculations in detail in Appendix B. It is easiest to calculate the sum of the probabilities like in the previous subsection.
We obtain the explicit expression:
\begin{equation}\label{prox}
\begin{split}
&\frac{1}{(N_c^2-1)^2}\sum_{a_l,a_k,a}\mathcal{P}^1_{N-1}(\{\mathbf{x}_i,a_i\}; \mathbf{x}_1,a_1...\mathbf{x}_{l-1},a_{l-1};\mathbf{x}_{l+1},a_{l+1}...\mathbf{x}_{k-1},a_{k-1,}\mathbf{x}_{k+1},a_{k+1}...\mathbf{x}_N,a_N; \mathbf{x},a)\\
%\mathcal{P}_{N-1}(\mathbf{x}_l, \mathbf{x}_k)  =
 &\ \ \ \ \ \ \ \ \ \ \ \ \ \ \ \ \ \ \ \ \ \ \ \ \ \ \ =-\frac{5}{24}\frac{N_c^4}{(N_c^2-1)}  \frac{\Delta}{2\pi}g^2 \color{black}
 \mathcal{B}_i(\mathbf{x}-\mathbf{x}_l) \mathcal{B}_i(\mathbf{x}-\mathbf{x}_k) \phi(\mathbf{x}-\mathbf{x}_l) \phi(\mathbf{x}-\mathbf{x}_k)\, .
 \end{split}
\end{equation}

We note the following. In principle eq.(\ref{prox}) with $\mathbf{x}=\mathbf{x}_k$ has to be added to eq.(\ref{pro0})  to form the total probability to annihilate a gluon at $\mathbf{x}_l$. However the expression in eq.(\ref{prox}) is $O(\alpha_s^{3})$, while eq.(\ref{pro0}) is $O(\alpha_s)$.  Being parametrically smaller, the contribution from eq.(\ref{prox}) can never compensate the negative contribution of eq.(\ref{pro0}). Thus we conclude that a state where one of the existing gluons disappears in one step of the evolution has a  negative probability.

 When $\mathbf{x}\ne\mathbf{x}_l,\mathbf{x}_k$, eq.(\ref{prox}) by itself constitutes a probability to find a state where two gluons have been annihilated and one new gluon created at a different transverse position. Examining the RHS of eq.(\ref{prox}) we see that its sign is not fixed but rather depends on the position of the point  $\mathbf{x}$. It is obvious for example, that at least for configurations where this extra  gluon is created far away from all the existing  gluons, i.e.  $|\mathbf{x}|\gg|\mathbf{x}_l|,|\mathbf{x}_k|$, this probability is also negative. %In this situation eq.(\ref{prox}) by itself indeed gives the complete probability of the state with gluons in this configuration, since the state is orthogonal to the one that contains only the original gluons.

We conclude that appearance of negative probabilities is ubiquitous in the JIMWLK evolution and thus the violation of unitarity is quite brazen.

\section{Discussion}

Let us recap the results of this paper. 

We have given the formal definition of the algorithm for calculation of scattering amplitudes in QCD RFT in terms of the ``correlators" of strings of $U$'s and $\bar U$'s - the basic RFT degrees of  freedom. As part of this algorithm we have also provided explicit realization of the field algebra  of RFT. This algebra has an intuitive interpretation in terms of QCD gluon-gluon scattering amplitude.

Starting from the eikonal approximation for  calculation of the QCD amplitudes, we  have formulated the unitarity conditions on the $H_{RFT}$. These conditions stem directly from the requirement that the RFT calculation has to be equivalent to a calculation performed in terms of normalized QCD wave functions. It lead us to identify certain coefficients in the action of $H_{RFT}$ on an RFT amplitude with probabilities, which thus have to be positive and bounded by unity.

We further discussed how these unitarity conditions are realized in the JIMWLK limit of RFT, where one of the scattering objects is dilute and the other one is dense. In this limit the RFT Hamiltonian - $H_{JIMWLK}$ is known. We found that when acting on the dilute projectile, the action of $H_{JIMWLK}$ indeed satisfies the unitarity conditions. On the other hand we have proven for the first time, that the unitarity is violated by $H_{JIMWLK}$ when acting on the dense target. In this case we have demonstrated that negative ``probabilities" arise. 

We note that the probabilities arising in the target evolution are not only negative but can also be infrared divergent. For example the integral in eq.(\ref{pro0}) logarithmically diverges at large values of $\mathbf{x}$. Similar infrared divergent probabilities arise in the evolution of a projectile, {\it if the projectile is not a color singlet state}. However for color singlet projectiles all probabilities are IR finite. On the other hand the IR divergence in eq.(\ref{pro0}) is independent of the global color representation of the target state. We believe this is another symptom of the inadequate treatment of the dense target in the JIMWLK approximation, and should disappear once the unitarity of the evolution is restored in a more refined approach.

The action of $H_{JIMWLK}$ on the dense target state bears many similarities to  the zero dimensional toy model studied in \cite{KLL}.  We observed that when interpreted in terms of the energy evolution of the underlying QCD state, 
apart from generating negative probabilities the JIMWLK evolution possesses another curious property. Generally one expects that the evolution of a QCD state in energy leads to increase in number of gluons. The process of physical gluon annihilation in the wave function of QCD is not associated with leading longitudinal logarithms, but emission of gluons is, and thus the number of gluons should always increase under the evolution. One does expect  the {\bf rate of growth} of number of gluons to behave differently for a dilute and a dense system. In a dilute system this rate is proportional to the number of gluons present in the wave function, as emissions from different color sources are independent. In a dense state on the other hand the rate of growth is constant and independent of the number of existing gluons, since the process of emission is fully coherent. It is in this sense that one talks about ``saturation" in the dense gluon state. In both situations however, the net number of gluons should grow with energy. However as discussed in detail in Section 5, JIMWLK evolution of the dense target leads to admixture in the initial state of states with {\bf smaller} number of gluons, and never larger. So rather than generating more gluons with positive probabilities, the evolution generates less gluons with negative probabilities! These two minuses make a plus, which results in the correct  evolution of the $S$-matrix when viewed from the target side.

It is amusing to note that the evolution of the dilute projectile is frequently referred to in the literature as ``gluon splitting", while that of the dense target as ``gluon merging".  The picture of the evolution in terms of the wave function we described above  conforms with this terminology in a peculiar way. Evolution of the projectile indeed is due to splitting of gluons in the wave function, while the dense target indeed experiences  gluon merging, since the number of gluons decreases with energy. This of course with the disclaimer  that the merging happens with negative probability and thus cannot be interpreted as a real physical process. We stress, that physically the gluons in QCD wave function do not merge, but always ``split", albeit the splitting process is coherent at high density.

Our analysis in this paper makes it clear what is the physical mechanism which leads to misidentification of the wave function evolution as ``merging". The root cause lies in limiting scattering amplitude of any target gluon by at most two gluon exchanges. Any target gluon that exchanges two gluons with the Hamiltonian does not scatter on the projectile, and thus effectively disappears from the wave function. We note that this situation again is very similar to the zero dimensional toy model. This observation also suggests a possible way forward to restore the unitarity: one should allow for arbitrary number of exchanges of a given target gluon with the projectile. This is necessary for regaining unitarity, and simultaneously is also required at high enough energies where the projectile is not dilute anymore. 

However even if one allows for more exchanges it is most likely that the unitarity will not be fully restored. This is the lesson we learned in a zero dimensional toy model. One also needs to modify the RFT Hamiltonian itself. In the perfect world this modification should be derived directly within QCD similarly to the derivation of $H_{JIMWLK}$. Some attempts in this direction have been made in the past \cite{foam,diamond,Balitsky05,GLV}. This is a hard problem and it is still awaiting solution.

A somewhat less ambitious approach could be to try and determine the full RFT Hamiltonian by requiring that it is  self dual. Although self duality may not be sufficient to restore unitarity, it is likely to be a necessary condition. One could try the effective field theory approach, i.e. given the degrees of freedom (in our case $U$ and $\bar U$) to search for a Hamiltonian which possesses the known symmetries. In our case the relevant symmetries are self duality and an additional pair of discrete $Z_2$ symmetries - the charge conjugation and the signature symmetry \cite{reggeon}. This is left for future work.

%One member of this family is very reminiscent of the ``diamond action" discussed in the literature several years ago. However despite the superficial similarity we were able to show that it is not identical to the "diamond action" and differs from it in some significant aspects. We note that our "bottom up" approach does not allow us to decide which one of the candidates we have found is the right one, and in fact whether any one of them is the correct QCD RFT Hamiltonian. At the minimum the correct Hamiltonian should satisfy the unitarity constraints. Unfortunately analyzing the unitarity conditions beyond the JIMWLK limit turned out to be a complicated problem which at this point we were not able to solve. We believe it is a very important question and are planning to address it in future work.

 %  %%%%%%%%%%%%%%%%%%%%%%%%%%%%%%%%%%%%%%%%%%%%%%%%%%
\section{Acknowledgements}
  %%%%%%%%%%%%%%%%%%%%%%%%%%%%%%%%%%%%%%%%%%% 
   We thank our colleagues at Tel Aviv university and UTFSM for
 encouraging discussions.AK and Ming Li were supported by the NSF Nuclear Theory grants 1614640 and 1913890.
  EL was supported  by 
 ANID PIA/APOYO AFB180002 (Chile) and  Fondecyt (Chile) grant  
\# 1180118.  
ML was supported by the Israeli Science Foundation (ISF) grant \#1635/16.
ML and AK were also supported by the Binational Science Foundation grants \#2015626, \#2018722, and the Horizon 2020 RISE
 "Heavy ion collisions: collectivity and precision in saturation physics"  under grant agreement No. 824093.
This work has been performed
in the framework of COST Action CA15213 ``Theory of hot matter and relativistic heavy-ion collisions" (THOR).

\appendix
\section{Color algebra for $N$ gluon state}
\subsection{Inverting the relation between $U$ and $\alpha$}
In this section, we explicitly verify eq. \eqref{eq:2alpha_to_U}.  We start with the color identity \cite{Haber:2019sgz}
\begin{equation}
\begin{split}
\mathrm{Tr}(t^at^qt^bt^p) &= \frac{1}{4N_c} \delta_{aq}\delta_{bp} + \frac{1}{8} (d_{aqe}d_{bpe} - f_{aqe}f_{bpe} + if_{aqe}d_{bpe} + if_{bpe}d_{aqe})\\
&= \frac{1}{4N_c} \delta_{aq}\delta_{bp}  + \frac{1}{8}\left( (D^qD^p)_{ab} - (T^qT^p)_{ab} + (T^q D^p)_{ab} -(D^qT^p)_{ab}\right)\\
\end{split}
\end{equation}
with the identifications $f_{abc} = iT^a_{bc}$ and $d_{abc}= D^a_{bc}$. Here $d_{abc}$ is a totally symmetric  tensor with respect to the indices $a, b,c$. One then obtains
\begin{equation}
\mathrm{Tr}(t^at^qt^bt^p) + (p\leftrightarrow q) =\frac{1}{4N_c} (\delta_{aq}\delta_{bp}  + \delta_{ap}\delta_{bq}) + \frac{1}{8} \left((D^qD^p+D^pD^q)_{ab}-(T^pT^q+T^qT^p)_{ab}\right)
\end{equation}
using the relation $(T^qD^p)_{ab} - (D^pT^q)_{ab} = if_{qpr} D^r_{ab}$. Note that the imaginary terms cancel.

We now use the expansion of $U^{ab}(\mathbf{x})$ to second order 
\begin{equation}
U^{ab}(\mathbf{x}) = \delta^{ab} + i gT^e_{ab} \alpha^e(\mathbf{x}) -\frac{g^2}{2} (T^mT^n)_{ab}\alpha^m(\mathbf{x})\alpha^n(\mathbf{x})
\end{equation}
%and explicitly demonstrate that the relation in eq. (\ref{uweak}) is the correct relation. 
and substitute it %the expression $U^{ab}(\mathbf{x})$
 into the right hand side of eq. \eqref{eq:2alpha_to_U}. We find that the zeroth order term in $\alpha$ vanishes after using $\mathrm{Tr} (D^pD^q) = \delta^{pq}(N_c^2-4)/N_c $ and $\mathrm{Tr}(T^pT^q) =N_c \delta^{pq}$. The first order term in $\alpha$ also vanishes because $T^e_{ab}$ is antisymmetric with respect to $a,b$. The second order terms combine into
\begin{equation}
\begin{split}
&-g^2\frac{1}{N_c}(T^mT^n)_{pq} \alpha^m(\mathbf{x}) \alpha^n(\mathbf{x}) - \frac{g^2}{4}\alpha^m(\mathbf{x}) \alpha^n(\mathbf{x}) \Big(\mathrm{Tr}[(D^qD^p+D^pD^q)T^nT^m]-\mathrm{Tr}[(T^pT^q+T^qT^p)T^nT^m]\Big)\\
=& g^2 \alpha^p(\mathbf{x}) \alpha^q(\mathbf{x})
\end{split}
\end{equation}
where we have used the identity
\begin{equation}
\mathrm{Tr}(T^aT^bT^cT^d) - \mathrm{Tr}(T^aT^b D^cD^d) = \delta_{ad}\delta_{bc}+\delta_{ac}\delta_{bd} + \frac{2}{N_c} (T^aT^b)_{dc}\, .
\end{equation}
Thus relation  \eqref{eq:2alpha_to_U} is  proved. 
%All the identities used can be found in the reference \cite{Haber:2019sgz}. We will also need the identity
%\begin{equation}
%\mathrm{Tr}(t^at^bt^at^c) = -\frac{1}{4N_c}\delta^{bc}
%\end{equation}

\subsection{The virtual term} 
In this section we provide details of the calculation leading to eq.(\ref{color}). 
In the following no summation over repeated indexes $a_l$ and $a$ is assumed.

For adjoint generators, one has the following identity
\begin{equation}
\mathrm{Tr}(T^qT^{a_l}T^{a_l}T^p) = \delta_{qp}\delta^{a_la_l} + \delta_{pa_l}\delta_{qa_l}\, .
\end{equation}
For fundamental generators, one has
\begin{equation}
\left[\mathrm{Tr}(t^at^qt^at^p) + \mathrm{Tr}(t^at^pt^at^q)\right] = \frac{1}{2N_c}\delta_{ap}\delta_{aq} + \frac{1}{8} \left[(D^pD^q+D^qD^p)_{aa} -(T^pT^q+T^qT^p)_{aa}\right].
\end{equation}
Consequently,
\begin{equation}
\begin{split}
&\mathrm{Tr}(T^qT^{a_l}T^{a_l}T^p) \left[\mathrm{Tr}(t^at^qt^at^p) + \mathrm{Tr}(t^at^pt^at^q)\right]\\
=&\frac{1}{2N_c} \delta^{aa}\delta^{a_la_l} + \frac{1}{4}\delta^{a_la_l} [(D^pD^p)_{aa} -(T^pT^p)_{aa}]\\
&+\frac{1}{2N_c} \delta_{aa_l}\delta_{aa_l} + \frac{1}{4} [(D^{a_l}D^{a_l})_{aa} -(T^{a_l}T^{a_l})_{aa}]\\
=&-\frac{1}{2N_c} \delta^{aa}\delta^{a_la_l}+\frac{1}{2N_c} \delta_{aa_l}\delta_{aa_l} + \frac{1}{4} [(D^{a_l}D^{a_l})_{aa} -(T^{a_l}T^{a_l})_{aa}]\\
\end{split}
\end{equation}
We have used $D^pD^p = (N_c^2-4)/N_c \mathbf{1}$ and $T^pT^p = N_c\mathbf{1}$.
Then
\begin{equation}
\begin{split}
&-\frac{1}{2}(T^{c_1}T^{c_2})_{pq}  (T^{c_1}T^{c_2}+T^{c_2}T^{c_1})_{a_la_l} 4 \left[\mathrm{Tr}(t^at^qt^at^p) + \mathrm{Tr}(t^at^pt^at^q)\right]\\
=&-2 \left[\mathrm{Tr}(T^qT^{a_l}T^{a_l}T^p) +\mathrm{Tr}(T^pT^{a_l}T^{a_l}T^q)\right] \left[\mathrm{Tr}(t^at^qt^at^p) + \mathrm{Tr}(t^at^pt^at^q)\right]\\
=&-4 \left(-\frac{1}{2N_c} \delta^{aa}\delta^{a_la_l}+\frac{1}{2N_c} \delta_{aa_l}\delta_{aa_l} + \frac{1}{4} [(D^{a_l}D^{a_l})_{aa} -(T^{a_l}T^{a_l})_{aa}]\right)\\
\end{split}
\end{equation}

If we set $a_l = a$, then the probability vanishes, since $d_{aac}=0$ and $f_{aac}=0$.
If we sum over index $a$, we get $2N_c \delta^{a_la_l}$.

\section{Calculation of the probability for $N-1$ gluon state.}
In this Appendix we give details of the calculation of the probability for the state component with $N-1$ gluons that leads to eq.(\ref{prox}). Here two gluons are annihilated by the action of derivatives  $\delta/\delta \rho^a_{\mathbf{x}}$ from $\bar{U}, M_L, M_R$ and one gluon is created by the factor $U$ in $\tilde H_{JIMWLK}$, whose expression is given by eqs.\eqref{eq:fundamental_to_adjoint}, \eqref{eq:2alpha_to_U}.

In the calculation of probability $U^{ab}(\mathbf{x})$ is set to  $\delta^{ab}$ and we can use eq. \eqref{eq:U_becomes_delta}.  Including $\delta^{pq}$ and using $M_R^{qd} = M_L^{dq}$, one obtains
\begin{equation}\label{ai}
\begin{split}
&\delta^{pq}\left[\partial_{\mathbf{x}}^2\bar{U}^{ed}(\mathbf{x})\right]M_L^{pe}\left[ T^a\delta/\delta\rho^a(\mathbf{x})\right]M_R^{qd}\left[ T^a\delta/\delta\rho^a(\mathbf{x})\right]\\
=&\partial_{\mathbf{x}}^2\bar{U}^{ed}(\mathbf{x}) \left(M^2_L[T^a\delta/\delta\rho^a(\mathbf{x})]\right)^{de}
\end{split}
\end{equation}
where
\begin{equation}
M^2_L[x] = 1- x + \frac{5}{12}x^2 - \frac{1}{12}x^3 + \frac{1}{240}x^4+ \frac{1}{720} x^5 - \frac{1}{6048} x^6 + \ldots
\end{equation}

We need all terms  in eq.(\ref{ai}) that contain four factors of $\delta/\delta \rho^c_{\mathbf{x}}$. Those are
\begin{equation}\label{eq:four_d_drho_terms}
\begin{split}
& \partial_{\mathbf{x}}^2 (T^{c_1}_{ed}\frac{\delta}{\delta \rho^{c_1}_{\mathbf{x}}})\left(-\frac{1}{12} (T^{c_2}T^{c_3}T^{c_4})_{de}\frac{\delta}{\delta \rho^{c_2}_{\mathbf{x}}}\frac{\delta}{\delta \rho^{c_3}_{\mathbf{x}}}\frac{\delta}{\delta \rho^{c_4}_{\mathbf{x}}}\right)\\
&+\partial_{\mathbf{x}}^2 \left(\frac{1}{2!}(T^{c_1}T^{c_2})_{ed}\frac{\delta}{\delta \rho^{c_1}_{\mathbf{x}}}\frac{\delta}{\delta \rho^{c_2}_{\mathbf{x}}}\right)\left(\frac{5}{12} (T^{c_3}T^{c_4})_{de}\frac{\delta}{\delta \rho^{c_3}_{\mathbf{x}}}\frac{\delta}{\delta \rho^{c_4}_{\mathbf{x}}}\right)\\
&+\partial_{\mathbf{x}}^2 \left(\frac{1}{3!}(T^{c_1}T^{c_2}T^{c_3})_{ed}\frac{\delta}{\delta \rho^{c_1}_{\mathbf{x}}}\frac{\delta}{\delta \rho^{c_2}_{\mathbf{x}}}\frac{\delta}{\delta \rho^{c_3}_{\mathbf{x}}}\right)\left(- (T^{c_4})_{de}\frac{\delta}{\delta \rho^{c_4}_{\mathbf{x}}}\right)\\
&+\partial_{\mathbf{x}}^2 \left(\frac{1}{4!}(T^{c_1}T^{c_2}T^{c_3}T^{c_4})_{ed}\frac{\delta}{\delta \rho^{c_1}_{\mathbf{x}}}\frac{\delta}{\delta \rho^{c_2}_{\mathbf{x}}}\frac{\delta}{\delta \rho^{c_3}_{\mathbf{x}}}\frac{\delta}{\delta \rho^{c_4}_{\mathbf{x}}}\right)\delta_{de}.\\
\end{split}
\end{equation}
Among them, the subset of terms that involve the spatial derivatives acting only on one of the $\delta/\delta \rho^c_{\mathbf{x}}$ add up to zero because
\begin{equation}
\begin{split}
&\mathrm{Tr} (T^{c_1}T^{c_2}T^{c_3}T^{c_4}) \left(\partial_{\mathbf{x}}^2 \frac{\delta}{\delta \rho^{c_1}_{\mathbf{x}}}\right)\frac{\delta}{\delta \rho^{c_2}_{\mathbf{x}}}\frac{\delta}{\delta \rho^{c_3}_{\mathbf{x}}}\frac{\delta}{\delta \rho^{c_4}_{\mathbf{x}}}\\
&\times \left( 1\times(-\frac{1}{12}) + 2\times \frac{1}{2!} \times \frac{5}{12} + 3\times\frac{1}{3!} \times (-1) + 4 \times \frac{1}{4!} \times 1 \right) =0.\\
\end{split}
\end{equation}
These terms would have contained $\phi(\mathbf{0})$, which is divergent, upon acting on the dense projectile if they did not vanish. Thus this is a demonstration that to this order our result does not depend on the constant in the definition of the potential $\phi$. 

Note that, when the two spatial derivatives only act on one of $\delta/\delta \rho^c_{\mathbf{x}}$, the following relation is valid
\begin{equation}
\begin{split}
&\partial_{\mathbf{x}}^2 \left(\frac{1}{m!} (T^{c_1} T^{c_2} \ldots T^{c_m}) \frac{\delta}{\delta \rho^{c_1}_{\mathbf{x}}}\frac{\delta}{\delta \rho^{c_2}_{\mathbf{x}}}\ldots \frac{\delta}{\delta \rho^{c_m}_{\mathbf{x}}}\right) \\
=& m\times \frac{1}{m!} \{T^{c_1} T^{c_2} \ldots T^{c_m}\}\left(\partial_{\mathbf{x}}^2\frac{\delta}{\delta \rho^{c_1}_{\mathbf{x}}}\right)\frac{\delta}{\delta \rho^{c_2}_{\mathbf{x}}}\ldots \frac{\delta}{\delta \rho^{c_m}_{\mathbf{x}}}.\\
\end{split}
\end{equation}
Here 
\begin{equation}\{T^{c_1} T^{c_2} \ldots T^{c_m}\}\equiv\frac{1}{m!}\sum_{P(1...m)}T^{c_{P_1}} T^{c_{P_2}} \ldots T^{c_{P_m}}
\end{equation} 
with the summation over all permutations of $c_1,...,c_m$.

When the two spatial derivatives act on two different  factors $\delta/\delta \rho^c_{\mathbf{x}}$, the following relation is valid
\begin{equation}
\begin{split}
&\partial_{\mathbf{x}}^2 \left(\frac{1}{m!} (T^{c_1} T^{c_2} \ldots T^{c_m}) \frac{\delta}{\delta \rho^{c_1}_{\mathbf{x}}}\frac{\delta}{\delta \rho^{c_2}_{\mathbf{x}}}\ldots \frac{\delta}{\delta \rho^{c_m}_{\mathbf{x}}}\right) \\
=& m(m-1) \times \frac{1}{m!} \{T^{c_1} T^{c_2} \ldots T^{c_m}\}\left(\partial_{\mathbf{x}}\frac{\delta}{\delta \rho^{c_1}_{\mathbf{x}}}\right)\left(\partial_{\mathbf{x}}\frac{\delta}{\delta \rho^{c_2}_{\mathbf{x}}}\right)\ldots \frac{\delta}{\delta \rho^{c_m}_{\mathbf{x}}}.\\
\end{split}
\end{equation}
Using these relations, eq. \eqref{eq:four_d_drho_terms} becomes 
\begin{equation}\label{eq:53}
\begin{split}
&\left[\frac{5}{12} \mathrm{Tr}(\{T^{c_1}T^{c_2}\} \{T^{c_3}T^{c_4}\}) -  \mathrm{Tr}(\{T^{c_1}T^{c_2}T^{c_3}\}T^{c_4}) + \frac{1}{2} \mathrm{Tr}(\{T^{c_1}T^{c_2}T^{c_3}T^{c_4}\})\right]\\
&\times \left(\partial_{\mathbf{x}}\frac{\delta}{\delta \rho^{c_1}_{\mathbf{x}}}\right)\left(\partial_{\mathbf{x}}\frac{\delta}{\delta \rho^{c_2}_{\mathbf{x}}}\right) \frac{\delta}{\delta \rho^{c_3}_{\mathbf{x}}}\frac{\delta}{\delta \rho^{c_4}_{\mathbf{x}}}\\
\end{split}
\end{equation}
With the understanding that the following traces are multiplied by functions symmetric with respect to interchange of $c_1$ and $ c_2$ and separately $c_3$ and $c_4$, we can identify them as
\begin{equation}
\begin{split}
&\mathrm{Tr}(\{T^{c_1}T^{c_2}\} \{T^{c_3}T^{c_4}\})  \rightarrow \mathrm{Tr} (T^{c_1}T^{c_2}T^{c_3}T^{c_4}),\\
&\mathrm{Tr}(\{T^{c_1}T^{c_2}T^{c_3}\}T^{c_4}) \rightarrow \frac{2}{3}\mathrm{Tr} (T^{c_1}T^{c_2}T^{c_3}T^{c_4}) + \frac{1}{3}\mathrm{Tr} (T^{c_1}T^{c_3}T^{c_2}T^{c_4}),\\
&\mathrm{Tr}(\{T^{c_1}T^{c_2}T^{c_3}T^{c_4}\})\rightarrow \frac{2}{3}\mathrm{Tr} (T^{c_1}T^{c_2}T^{c_3}T^{c_4}) + \frac{1}{3}\mathrm{Tr} (T^{c_1}T^{c_3}T^{c_2}T^{c_4})\, .\\
\end{split}
\end{equation}
Moreover we have
\begin{equation}
\begin{split}
& \left(\partial_{\mathbf{x}}\frac{\delta}{\delta \rho^{c_1}_{\mathbf{x}}}\right)\left(\partial_{\mathbf{x}}\frac{\delta}{\delta \rho^{c_2}_{\mathbf{x}}}\right) \frac{\delta}{\delta \rho^{c_3}_{\mathbf{x}}}\frac{\delta}{\delta \rho^{c_4}_{\mathbf{x}}} U^{a_lb_l}(\mathbf{x}_l) U^{a_kb_k}(\mathbf{x}_k)\\
=&\frac{g^4}{4} (T^{c_1}T^{c_2} + T^{c_2}T^{c_1})_{a_lb_l}(T^{c_3}T^{c_4} + T^{c_4}T^{c_3})_{a_kb_k} [\mathcal{B}_i(\mathbf{x}-\mathbf{x}_l)]^2 [\phi(\mathbf{x}-\mathbf{x}_k)]^2\\
&+\frac{g^4}{4} \left[(T^{c_1}T^{c_3} + T^{c_3}T^{c_1})_{a_lb_l}(T^{c_2}T^{c_4} + T^{c_4}T^{c_2})_{a_kb_k} +(T^{c_1}T^{c_4} + T^{c_4}T^{c_1})_{a_lb_l}(T^{c_2}T^{c_3} + T^{c_3}T^{c_2})_{a_kb_k}\right]\\
&\times \mathcal{B}_i(\mathbf{x}-\mathbf{x}_l) \mathcal{B}_i(\mathbf{x}-\mathbf{x}_k) \phi(\mathbf{x}-\mathbf{x}_l) \phi(\mathbf{x}-\mathbf{x}_k)\\
&+(l \leftrightarrow k).\\
\end{split}
\end{equation}
To make the calculation manageable we will
%At the end, we want to look at the diagonal terms by setting $a_l=b_l$ and $a_k=b_k$ , which represents the forward scatterings. If these probabilities are positive, summation over $a_l$ and $a_k$ would also lead to positive probabilities. On the othe hand, if summation of probabilities turns out to be negative or zero, it means some of the probabilities are negative, thus violating the unitarity.  We therefore consider 
take the traces over $a_l, b_l$ and $a_k, b_k$. 
\begin{equation}
\begin{split}
& \left(\partial_{\mathbf{x}}\frac{\delta}{\delta \rho^{c_1}_{\mathbf{x}}}\right)\left(\partial_{\mathbf{x}}\frac{\delta}{\delta \rho^{c_2}_{\mathbf{x}}}\right) \frac{\delta}{\delta \rho^{c_3}_{\mathbf{x}}}\frac{\delta}{\delta \rho^{c_4}_{\mathbf{x}}}\frac{1}{(N_c^2-1)}\mathrm{Tr}[ U^{a_lb_l}(\mathbf{x}_l)]\frac{1}{(N_c^2-1)} \mathrm{Tr}[ U^{a_kb_k}(\mathbf{x}_k)]\\
=&\frac{g^4}{2} N_c^2 \delta^{c_1c_2}\delta^{c_3c_4} [\mathcal{B}_i(\mathbf{x}-\mathbf{x}_l)]^2 [\phi(\mathbf{x}-\mathbf{x}_l)]^2+ \frac{g^4}{2}N_c^2(\delta^{c_1c_3}\delta^{c_2c_4} + \delta^{c_1c_4}\delta^{c_2c_3}) \\
&\quad \times \mathcal{B}_i(\mathbf{x}-\mathbf{x}_l) \mathcal{B}_i(\mathbf{x}-\mathbf{x}_k) \phi(\mathbf{x}-\mathbf{x}_l) \phi(\mathbf{x}-\mathbf{x}_k) \frac{1}{(N_c^2-1)^2}
\end{split}
\end{equation}
From Eq. \eqref{eq:53}, one notes that
\begin{equation}
\left(\frac{1}{12}\mathrm{Tr} (T^{c_1}T^{c_2}T^{c_3}T^{c_4}) - \frac{1}{6} \mathrm{Tr} (T^{c_1}T^{c_3}T^{c_2}T^{c_4})\right) \delta^{c_1c_2}\delta^{c_3c_4} =0\, ,
\end{equation}
following from the identities $T^aT^a = N_c \mathbf{1}$, $\mathrm{Tr}(T^aT^bT^aT^c) = \frac{1}{2}N_c^2 \delta^{bc}$, $\mathrm{Tr}(\mathbf{1}) = N_c^2-1$.
Furthermore 
\begin{equation}
\left(\frac{1}{12}\mathrm{Tr} (T^{c_1}T^{c_2}T^{c_3}T^{c_4}) - \frac{1}{6} \mathrm{Tr} (T^{c_1}T^{c_3}T^{c_2}T^{c_4})\right) (\delta^{c_1c_3}\delta^{c_2c_4} + \delta^{c_1c_4}\delta^{c_2c_3}) = -\frac{5}{24} N_c^2(N_c^2-1)\, .
\end{equation}
Then the summed probability is
\begin{equation}
\mathcal{P}_{N-1}(\mathbf{x}_l, \mathbf{x}_k)  = -\frac{5}{24}\frac{N_c^4}{(N_c^2-1)} g^4 \mathcal{B}_i(\mathbf{x}-\mathbf{x}_l) \mathcal{B}_i(\mathbf{x}-\mathbf{x}_k) \phi(\mathbf{x}-\mathbf{x}_l) \phi(\mathbf{x}-\mathbf{x}_k)\, .
\end{equation}

%\section*{Acknowledgements}

%AK and Ming Li were supported by the NSF Nuclear Theory grants 1614640 and 1913890.
%ML was supported by the Israeli Science Foundation (ISF) grant \#1635/16.
%ML and AK were also supported by the Binational Science Foundation grants \#2015626, \#2018722, and the Horizon 2020 RISE
 %"Heavy ion collisions: collectivity and precision in saturation physics"  under grant agreement No. 824093.
%This work has been performed
%in the framework of COST Action CA15213 ``Theory of hot matter and relativistic heavy-ion collisions" (THOR).


\newpage

\begin{thebibliography}{99}
 
  \bibitem{KLL} A. Kovner, E.Levin and M. Lublinsky, JHEP {\bf 1608} (2016) 031
  

\bibitem{gribov} V.~N.~Gribov,
  %``A Reggeon Diagram Technique,''
  Sov.\ Phys.\ JETP {\bf 26}, 414 (1968)
  [Zh.\ Eksp.\ Teor.\ Fiz.\  {\bf 53}, 654 (1967)].
  %%CITATION = SPHJA,26,414;%%


\bibitem{BFKL}
 E. A. Kuraev, L. N. Lipatov, and F. S. Fadin, { Sov. Phys.
JETP}
                {\bf 45}, 199 (1977); \,\,\,
Ya. Ya. Balitsky and L. N. Lipatov,
               { Sov. J. Nucl. Phys.}\, {\bf 28}, 22 (1978).
               
               
\bibitem{glr} L. Gribov, E. Levin and M. Ryskin, Phys. Rept. {\bf 100}, 1, 1983.

\bibitem{MUPA}
A. H. Mueller and J. Qiu, 
Nucl. Phys. {\bf B268} (1986) 427;\, H. Mueller and B. Patel,
Nucl. Phys. {\bf B425} (1994) 471.

\bibitem{MUDI}
  A.~H.~Mueller,
  %``Soft gluons in the infinite momentum wave function and the BFKL pomeron,''
  Nucl.\ Phys.\ B {\bf 415} (1994) 373;\,\,\,
  %%CITATION = doi:10.1016/0550-3213(94)90116-3;%%
  %``Unitarity and the BFKL pomeron,''
  Nucl.\ Phys.\ B {\bf 437} (1995) 107;\\
%  [hep-ph/9408245];\\
  %%CITATION = doi:10.1016/0550-3213(94)00480-3;%%
  %456 citations counted in INSPIRE as of 25 Apr 2016  
      A.~H.~Mueller and B.~Patel, Nucl. Phys. B {\bf 425}, 471, 1994.         
               
   \bibitem{LIREV}
                L.~N.~Lipatov,
  %``Small x physics in perturbative QCD,''
  Phys.\ Rept.\  {\bf 286} (1997) 131.
 % [hep-ph/9610276].
  %%CITATION = doi:10.1016/S0370-1573(96)00045-2;%%               
               
\bibitem{LipatovFT}
  L.~N.~Lipatov,
  %``High-energy scattering in QCD and in quantum gravity and two-dimensional
  %field theories,''
  Nucl.\ Phys.\ B {\bf 365}, 614 (1991),
  %%CITATION = NUPHA,B365,614;%%
%``Gauge invariant effective action for high-energy processes in QCD,''
  Nucl.\ Phys.\ B {\bf 452}, 369 (1995), \\
 % [arXiv:hep-ph/9502308];\\
  %%CITATION = HEP-PH 9502308;%%
R.~Kirschner, L.~N.~Lipatov and L.~Szymanowski,
  %``Effective action for multi - Regge processes in QCD,''
  Nucl.\ Phys.\ B {\bf 425}, 579 (1994),
 % [arXiv:hep-th/9402010].
  %%CITATION = HEP-TH 9402010;%%
%``Symmetry properties of the effective action for high-energy scattering in
  %QCD,''
  Phys.\ Rev.\ D {\bf 51}, 838 (1995).
 % [arXiv:hep-th/9403082].
  %%CITATION = HEP-TH 9403082;%%


  \bibitem{bartels} J.~Bartels, Z.Phys. {\bf C60}, 471 (1993);\\
J.~Bartels and M.~Wusthoff, Z. Phys. {\bf C66}, 157 (1995);\,\,\,
 J.~Bartels and C.~Ewerz, JHEP  {\bf 9909}, 026 (1999), \\
 %hep-ph/9908454;\,\,\,
   C.~Ewerz,
  %``Reggeization in high-energy QCD,''
  JHEP {\bf 0104} (2001) 031.
%  [hep-ph/0103260].\,\,\,
  %%CITATION = doi:10.1088/1126-6708/2001/04/031;%%

\bibitem{BKP}
J.~Bartels,
% ``High-Energy Behavior In A Nonabelian Gauge Theory. 2. First 
%Corrections To
%T(N--->M) Beyond The Leading Lns Approximation,''
%
Nucl.\ Phys.\  {\bf B175}, 365 (1980);\\
%%CITATION = NUPHA,B175,365;%%
J.~Kwiecinski and M.~Praszalowicz,
% ``Three Gluon Integral Equation And Odd C Singlet Regge Singularities In
%QCD,''
%
Phys.\ Lett.\  {\bf B94}, 413 (1980).
%%CITATION = PHLTA,B94,413;%%

  
  \bibitem{mv} L.~McLerran and R.~Venugopalan, Phys. Rev. D{ \bf 49}: 2233-2241, (1994);  Phys. Rev. D {\bf49}: 3352-3355, (1994). 

\bibitem{Salam}
A.~H.~Mueller and G.~P.~Salam,
  %``Large multiplicity fluctuations and saturation effects in onium collisions,''
  Nucl.\ Phys.\ B {\bf 475}, 293 (1996);\\
%  [hep-ph/9605302];\\ 
  G.~P.~Salam,
  %``Studies of unitarity at small x using the dipole formulation,''
  Nucl.\ Phys.\ B {\bf 461}, 512 (1996).
%   [hep-ph/9509353].
  
   \bibitem{KOLE}
  Y.~V.~Kovchegov and E.~Levin,
  %``Diffractive dissociation including multiple pomeron exchanges in high parton density QCD,''
  Nucl.\ Phys.\ B {\bf 577} (2000) 221.
%  [hep-ph/9911523].
  %%CITATION = doi:10.1016/S0550-3213(00)00125-5;%%

  
  
  
   \bibitem{BRN}
M. A. Braun,
Eur. Phys. J. {\bf C16} (2000) 337;\\
% [arXiv:hep-ph/0001268]; \\
M. A. Braun and G. P. Vacca,
Eur. Phys. J. {\bf C6} (1999) 147; \\
%[arXiv:hep-ph/9711486];\\
J.~Bartels, M.~Braun and G.~P.~Vacca,
  %``Pomeron vertices in perturbative QCD in diffractive scattering,''
  Eur.\ Phys.\ J.\ C {\bf 40}, 419 (2005). 
  % [arXiv:hep-ph/0412218].
  %%CITATION = HEP-PH 0412218;%%
\\
 J.~Bartels, L.~N.~Lipatov and G.~P.~Vacca,
  %``Interactions of Reggeized gluons in the Moebius representation,''
  Nucl.\ Phys.\ B {\bf 706}, 391 (2005).
  %  [arXiv:hep-ph/0404110].

\bibitem{braun}  M.~A.~Braun,
 % {\it ``Nucleus-nucleus scattering in perturbative QCD with  $N_c \to  \infty$}
  Phys.\ Lett.\ B {\bf 483}, 115 (2000), 
  %[hep-ph/0003004];%\,\,\,{\it ``Nucleus nucleus interaction in the perturbative QCD,''}
  Eur.\ Phys.\ J.\ C {\bf 33}, 113 (2004);
   %[hep-ph/0309293];%\,\,\,{\it ``Conformal invariant pomeron interaction in the perurbative QCD with large $N_c$,''}
  Phys.\ Lett.\ B {\bf 632}, 297 (2006).  
  
 
\bibitem{BK}
I.~Balitsky,
%[arXiv:hep-ph/9509348];\,\,
%%CITATION = HEP-PH 9509348;%%
{Phys.\ Rev.} {\bf D60}, 014020 (1999);
%[arXiv:hep-ph/9812311];\,\,\,\,
%%CITATION = HEP-PH 9812311;%%
Y.~V.~Kovchegov,
{Phys.\ Rev.}  {\bf D60}, 034008  (1999).
%[arXiv:hep-ph/9901281].
%%CITATION = HEP-PH 9901281;%%
 
  
   \bibitem{reggeon} A. Kovner and M. Lublinsky, JHEP  \textbf{02} (2007), 058.
    % hep-ph/0512316.          
 
\bibitem{jimwlk} J. Jalilian Marian, A. Kovner, A. Leonidov and H. Weigert,
{Nucl. Phys.}{\bf  B504} 415 (1997);  
%e-Print   hep-ph/9701284
{ Phys. Rev.} {\bf D59} 014014 (1999);  \\
%e-Print   hep-ph/9706377
J. Jalilian Marian, A. Kovner and H. Weigert, { Phys. Rev.} {\bf D59} 
014015 (1999); \\
%   hep-ph/9709432;\\
A. Kovner and J.G. Milhano, {Phys. Rev.} {\bf D61} 014012 (2000); \\
%  hep-ph/9904420;\\
 A. Kovner, J.G. Milhano and H. Weigert,
{ Phys. Rev.} {\bf D62} 114005 (2000); \\
 H. Weigert, { Nucl.Phys.} {\bf A 703} (2002) 823.
 
 \bibitem{cgc}  E.Iancu, A. Leonidov and L. McLerran, {Nucl. Phys.} 
{\bf A 692} (2001) 583; {Phys. Lett.} {\bf B
510} (2001) 133;\\
E. Ferreiro, E. Iancu, A. Leonidov, L. McLerran;  
{Nucl. Phys.}{\bf A703} (2002) 489.
  
\bibitem{gluon}
  L.~N.~Lipatov,
  %``Reggeization Of The Vector Meson And The Vacuum Singularity In Nonabelian
  %Gauge Theories,''
  Sov.\ J.\ Nucl.\ Phys.\  {\bf 23}, 338 (1976)
  [Yad.\ Fiz.\  {\bf 23}, 642 (1976)];\\
  %%CITATION = SJNCA,23,338;%%
 L.~L.~Frankfurt and V.~E.~Sherman,
  %``Reggeization Of Vector Meson And Vacuum Singularity In Renormalizable
  %Yang-Mills Models,''
  Sov.\ J.\ Nucl.\ Phys.\  {\bf 23} (1976) 581;\\
  %%CITATION = SJNCA,23,581;%%
V.~S.~Fadin, M.~I.~Kotsky and R.~Fiore,
  %``Gluon Reggeization in QCD in the next-to-leading order,''
  Phys.\ Lett.\ B {\bf 359} (1995) 181.
  %%CITATION = PHLTA,B359,181;%%

\bibitem{motyka} S. Bondarenko and L. Motyka,  Phys. Rev. D {\bf 75} (2007) 114015.
%  hep-ph/0605185. 

  \bibitem{brauntarasov}  	
%BFKL pomeron propagator in the external field of the nucleus
M.A. Braun and  A. Tarasov; Nucl. Phys. B {\bf 851} (2011) 533-550;
 %1103.1747 [hep-ph];  
 Nucl. Phys. B {\bf 863} (2012) 495-509.
 %  1204.6066 [hep-ph]

  \bibitem{KLremark2} 
  A.~Kovner and M.~Lublinsky,
  %``More remarks on high energy evolution,''
  Nucl.\ Phys.\ A {\bf 767} 171 (2006).
 % [hep-ph/0510047].
  %%CITATION = doi:10.1016/j.nuclphysa.2005.12.010;%%

 
 

  
  %\bibitem{KLremark2} 
  %A.~Kovner and M.~Lublinsky,
  %``More remarks on high energy evolution,''
  %Nucl.\ Phys.\ A {\bf 767} 171 (2006).
 % [hep-ph/0510047].
  %%CITATION = doi:10.1016/j.nuclphysa.2005.12.010;%%

  
  \bibitem{ACJ}
  D.~Amati, L.~Caneschi and R.~Jengo,
  %``Summing Pomeron Trees,''
  Nucl.\ Phys.\ B {\bf 101} (1975) 397.
  %%CITATION = doi:10.1016/0550-3213(75)90604-5;%%
  %56 citations counted in INSPIRE as of 25 Apr 2016   
  
  \bibitem{AAJ}
  V.~Alessandrini, D.~Amati and R.~Jengo,
  %``One-Dimensional Quantum Theory of the Pomeron,''
  Nucl.\ Phys.\ B {\bf 108} (1976) 425.

  %%CITATION = doi:10.1016/0550-3213(76)90288-1;%%
  %42 citations counted in INSPIRE as of 25 Apr 2016
  \bibitem{JEN}
   R.~Jengo,
  %``Zero Slope Limit of the Pomeron Field Theory,''
  Nucl.\ Phys.\ B {\bf 108} (1976) 447.
  %%CITATION = doi:10.1016/0550-3213(76)90289-3;%%
  %35 citations counted in INSPIRE as of 25 Apr 2016  
    \bibitem{ABMC}
   D.~Amati, M.~Le Bellac, G.~Marchesini and M.~Ciafaloni,
  %``Reggeon Field Theory for alpha (o)>1,''
  Nucl.\ Phys.\ B {\bf 112} (1976) 107.
  %%CITATION = doi:10.1016/0550-3213(76)90492-2;%%
  %104 citations counted in INSPIRE as of 25 Apr 2016
\bibitem{CLR}
 M.~Ciafaloni, M.~Le Bellac and G.~C.~Rossi,
  %``Reggeon Quantum Mechanics: A Critical Discussion,''
  Nucl.\ Phys.\ B {\bf 130} (1977) 388.
  %%CITATION = doi:10.1016/0550-3213(77)90249-8;%%
  %31 citations counted in INSPIRE as of 25 Apr 2016
  
  \bibitem{CIAF}
    M.~Ciafaloni,
     %``Instanton Contributions in Reggeon Quantum Mechanics,''
  Nucl.\ Phys.\ B {\bf 146} (1978) 427.
  %%CITATION = doi:10.1016/0550-3213(78)90076-7;%%
  %11 citations counted in INSPIRE as of 25 Apr 2016
  
  \bibitem{KOZLE}
M.~Kozlov and E.~Levin,
%``Solution for the BFKL Pomeron Calculus in zero transverse dimensions,''
Nucl. Phys. A \textbf{779} (2006), 142.
%doi:10.1016/j.nuclphysa.2006.08.011
%[arXiv:hep-ph/0604039 [hep-ph]].
%39 citations counted in INSPIRE as of 14 Jun 2020  
  
  \bibitem{SHXI}  
    A.~I.~Shoshi and B.~W.~Xiao,
  %``Pomeron loops in zero transverse dimensions,''
  Phys.\ Rev.\ D {\bf 73} (2006) 094014.
%  [hep-ph/0512206].
  %%CITATION = doi:10.1103/PhysRevD.73.094014;%%
  %27 citations counted in INSPIRE as of 25 Apr 2016 
  
  \bibitem{BIT}
   J.-P.~Blaizot, E.~Iancu and D.~N.~Triantafyllopoulos,
  %``A Zero-dimensional model for high-energy scattering in QCD,''
  Nucl.\ Phys.\ A {\bf 784} (2007) 227.
  %[hep-ph/0606253].
  %%CITATION = doi:10.1016/j.nuclphysa.2006.11.127;%%
  %22 citations counted in INSPIRE as of 25 Apr 2016  
  	
\bibitem{nestor} N. Armesto, S. Bondarenko, J. G. Milhano and P. Quiroga, JHEP {\bf 0805} (2008) 103.
% arXiv:0803.0820 [hep-ph]. 
  
\bibitem{LEPRI}
  E.~Levin and A.~Prygarin,
  %``The BFKL Pomeron Calculus in zero transverse dimension: Summation of the Pomeron loops and the generating functional for the multiparticle production processes,''
  Eur.\ Phys.\ J.\ C {\bf 53} (2008) 385.
%  [hep-ph/0701178].
  %%CITATION = doi:10.1140/epjc/s10052-007-0458-5;%%
  %24 citations counted in INSPIRE as of 25 Apr 2016


 
\bibitem{KLduality} 
  A.~Kovner and M.~Lublinsky,
  %``From target to projectile and back again: Selfduality of high energy evolution,''
  Phys.\ Rev.\ Lett.\  {\bf 94}, 181603 (2005).
%  [hep-ph/0502119].
  %%CITATION = doi:10.1103/PhysRevLett.94.181603;%%


  
  \bibitem{yin} 
%The Yin and Yang of high energy chromodynamics: Scattering in black and white
A. Kovner and M. Lublinsky, Nucl.Phys. A{\bf 779} (2006) 220-243.
%e-Print: hep-ph/0604085  

 \bibitem{yinyang} A. Kovner and M. Lublinsky,  Phys.Rev. D{\bf 72} (2005) 074023.   %Nucl. Phys. A 779:220-243, (2006).
 %  hep-ph/0604085
 
 \bibitem{kovchegov}
%Quantum structure of the nonAbelian Weizsacker-Williams field for a very large nucleus
Y. V. Kovchegov, Phys.Rev. D{\bf 55} (1997) 5445-5455. 
%e-Print: hep-ph/9701229  

   \bibitem{ddd} A. Kovner and M. Lublinsky,  
  %Dense-dilute duality at work: dipoles of the target"
  Phys.Rev. D {\bf72} (2005) 074023;
%e-Print: hep-ph/05031
 
 %\bibitem{likovner} M. Li and A. Kovner, JHEP 05 (2020) 036,  e-Print: arXiv:2002.02282 
 
 
  
\bibitem{foam} A. Kovner, M. Lublinsky and U. Wiedemann, JHEP {\bf 0706}, 075 (2007); \\
%  0705.1713 [hep-ph];\\
T.~Altinoluk, A.~Kovner, M.~Lublinsky and J.~Peressutti,
  %``QCD Reggeon Field Theory for every day: Pomeron loops included,''
  JHEP {\bf 0903}, 109 (2009).
%e-Print: arXiv:0901.2559 [hep-ph].
  %%CITATION = doi:10.1088/1126-6708/2009/03/109;%%

 
    
  \bibitem{diamond} 
Y. Hatta, E. Iancu, L. McLerran, A. Stasto and D.N. Triantafyllopoulos, Nucl.Phys. A{\bf 764} (2006) 423, 
% e-Print: hep-ph/0504182 
  
  
  
  
\bibitem{Balitsky05}
  I.~Balitsky,
  Phys.\ Rev.\ D {\bf 72}, 074027 (2005).
%  ``High-enegy effective action from scattering of QCD shock waves,''
%  e-Print: arXiv:hep-ph/0507237.
  %%CITATION = HEP-PH 0507237;%%


\bibitem{GLV} 
  F.~Gelis, T.~Lappi and R.~Venugopalan,
  %``High energy factorization in nucleus-nucleus collisions,''
  Phys.\ Rev.\ D {\bf 78}, 054019 (2008).
%  [arXiv:0804.2630 [hep-ph]].
  %%CITATION = doi:10.1103/PhysRevD.78.054019;%%
 
\bibitem{Haber:2019sgz} 
  H.~E.~Haber,
 % ``\textit{Useful relations among the generators in the defining and adjoint representations of $SU(N)$},''
  arXiv:1912.13302 [math-ph]. 
  
  
  
  


\end{thebibliography}
\end{document}